\documentclass[10pt,conference]{IEEEtran}
\usepackage{cite}
\usepackage{amsmath,amssymb,amsfonts}
\usepackage{algorithmic}
\usepackage{graphicx}
\usepackage{textcomp}
\usepackage{xcolor}
\usepackage[hyphens]{url}
\usepackage{fancyhdr}
\usepackage{hyperref}

\definecolor{myyellow}{HTML}{FCFD99}

\usepackage{soul}
\definecolor{claudebg}{HTML}{F4F3EC}
\definecolor{claudesel}{HTML}{BA5B37}
\definecolor{claudebtn}{HTML}{BC5C3A}

\newcommand{\specialcell}[2][c]{%
\begin{tabular}[#1]{@{}c@{}}#2\end{tabular}}

\newcommand{\specialleftcell}[2][l]{%
\begin{tabular}[#1]{@{}l@{}}#2\end{tabular}}

\usepackage[normalem]{ulem}

\newcommand\entropyremove{\bgroup\markoverwith{\textcolor{violet}{\rule[0.5ex]{2pt}{1.0pt}}}\ULon}

\newcommand*\circled[1]{\tikz[baseline=(char.base)]{
            \node[shape=circle,draw,inner sep=0.2pt] (char) {#1};}}

\usepackage{booktabs}

\usepackage{pifont}

\usepackage{tikz}

\newcommand{\paraobserv}[1]{\vspace{0.5em}\noindent\textbf{#1}\vspace{0.5em}}

\usepackage[most]{tcolorbox}
\newtcolorbox{highlighted}{colback=yellow,coltext=red,breakable}

\newcommand{\hpcayear}{2026}

\newcommand{\hpcasubmissionnumber}{0}
\title{A Scalable Architecture for Efficient Multi-bit Fully Homomorphic Encryption}

\def\hpcacameraready{} 
\newcommand\hpcaauthors{Jiaao Ma$\dagger$, Ceyu Xu$\dagger$, and Lisa Wu Wills$\dagger$}
\newcommand\hpcaaffiliation{Duke University$\dagger$}
\newcommand\hpcaemail{jiaao.ma@duke.edu, ceyu.xu@duke.edu, lisa@cs.duke.edu}

\author{
  \ifdefined\hpcacameraready
    \IEEEauthorblockN{\hpcaauthors{}}
      \IEEEauthorblockA{
        \hpcaaffiliation{} \\
        \hpcaemail{}
      }
  \else
    \IEEEauthorblockN{\normalsize{HPCA \hpcayear{} Submission
      \textbf{\#\hpcasubmissionnumber{}}} \\
      \IEEEauthorblockA{
        Confidential Draft \\
        Do NOT Distribute!!
      }
    }
  \fi 
}

\fancypagestyle{camerareadyfirstpage}{%
  \fancyhead{}
  \renewcommand{\headrulewidth}{0pt}
  \fancyhead[C]{
    \ifdefined\aeopen
    \parbox[][12mm][t]{13.5cm}{\hpcayear{} IEEE International Symposium on High-Performance Computer Architecture (HPCA)}    
    \else
      \ifdefined\aereviewed
      \parbox[][12mm][t]{13.5cm}{\hpcayear{} IEEE International Symposium on High-Performance Computer Architecture (HPCA)}
      \else
      \ifdefined\aereproduced
      \parbox[][12mm][t]{13.5cm}{\hpcayear{} IEEE International Symposium on High-Performance Computer Architecture (HPCA)}
      \else
      \parbox[][0mm][t]{13.5cm}{\hpcayear{} IEEE International Symposium on High-Performance Computer Architecture (HPCA)}
    \fi 
    \fi 
    \fi 
    \ifdefined\aeopen 
      \includegraphics[width=12mm,height=12mm]{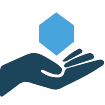}
    \fi 
    \ifdefined\aereviewed
      \includegraphics[width=12mm,height=12mm]{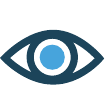}
    \fi 
    \ifdefined\aereproduced
      \includegraphics[width=12mm,height=12mm]{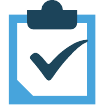}
    \fi
  }
  \fancyfoot[C]{}
}
\renewcommand{\headrulewidth}{0pt}

\begin{document}
\maketitle

\ifdefined\hpcacameraready 
  \thispagestyle{camerareadyfirstpage}
  \pagestyle{empty}
\else
  \thispagestyle{plain}
  \pagestyle{plain}
\fi

\newcommand{\hpcaheight}{0mm}
\ifdefined\eaopen
\renewcommand{\hpcaheight}{12mm}
\fi

\ifdefined\hpcacameraready
\fancypagestyle{camerareadyfirstpage}{%
  \fancyhead{}
  \renewcommand{\headrulewidth}{0pt}
  \fancyfoot[C]{}
}
\thispagestyle{plain}
\pagestyle{plain}
\fi


\begin{abstract}

  In the era of cloud computing, privacy-preserving computation offloading is crucial for safeguarding sensitive data.
Fully Homomorphic Encryption (FHE) enables secure processing of encrypted data, but the inherent computational complexity of FHE operations introduces significant computational overhead on the server side.
FHE schemes often face a tradeoff between efficiency and versatility.
While the CKKS scheme is highly efficient for polynomial operations, it lacks the flexibility of the binary TFHE (Torus-FHE) scheme, which offers greater versatility but at the cost of efficiency.
The recent multi-bit TFHE extension offers greater flexibility and performance by supporting native non-polynomial operations and efficient integer processing.
However, current implementations of multi-bit TFHE are constrained by its narrower numeric representation, which prevents its adoption in applications requiring wider numeric representations.

To address this challenge, we introduce Taurus, a hardware accelerator designed to enhance the efficiency of multi-bit TFHE computations.
Taurus supports ciphertexts up to 10 bits by leveraging novel FFT units and optimizing memory bandwidth through key reuse strategies.
We also propose a compiler with operation deduplication to improve memory utilization.
Our experiment results demonstrate that Taurus achieves up to $2600\times$ speedup over a CPU, $1200\times$ speedup over a GPU, and up to $7\times$ faster compared to the previous state-of-the-art TFHE accelerator.
Moreover, Taurus is the first accelerator to demonstrate privacy-preserving inference with large language models such as GPT-2.
These advancements enable more practical and scalable applications of privacy-preserving computation in cloud environments.

\end{abstract}

\section{Introduction}

In a world where data privacy concerns are increasingly significant, particularly with the rise of cloud-based services, there is a critical need to enable efficient privacy-preserving computation offloading.

\begin{figure}
    \centering
    \includegraphics[width=\linewidth]{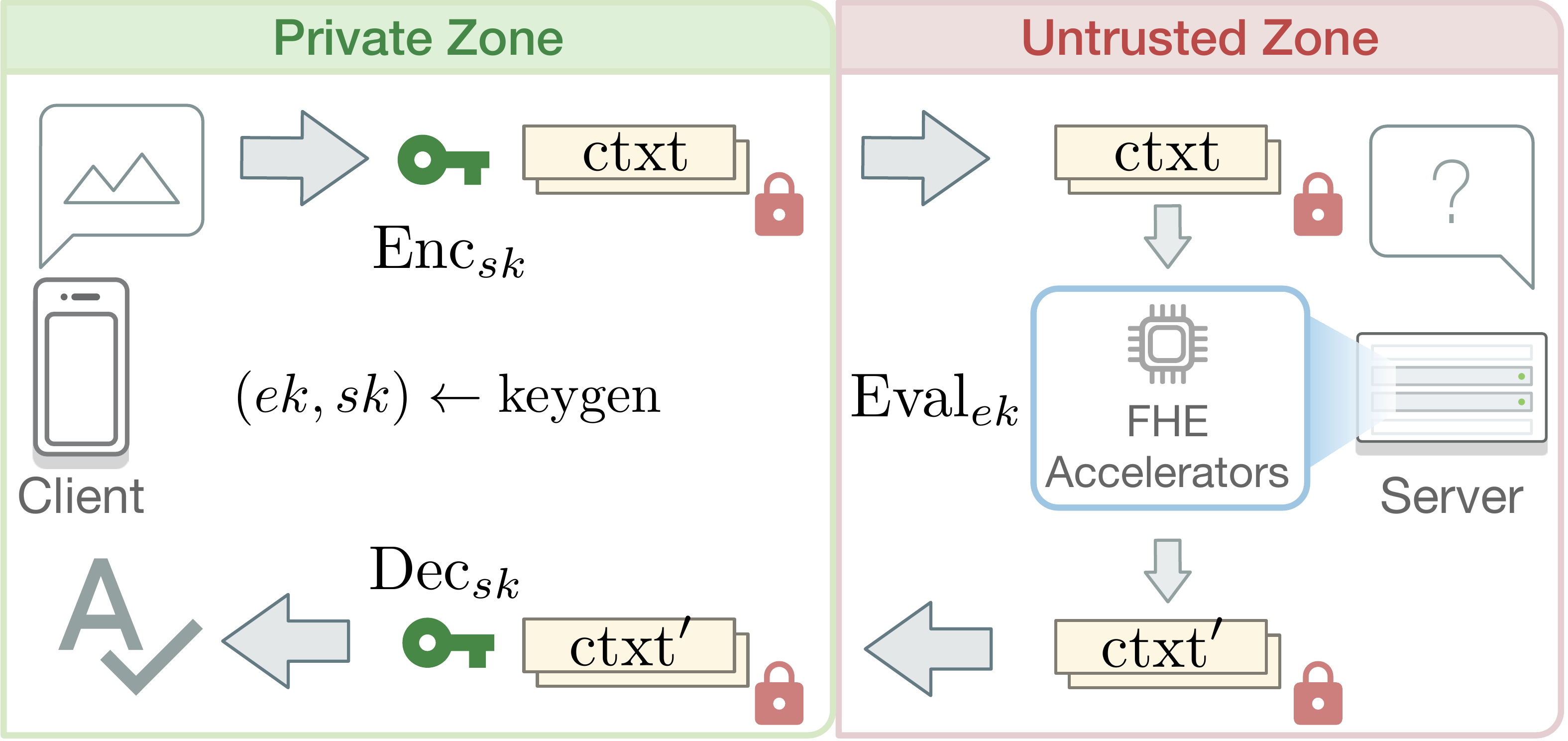}
    \caption{Secure Computation Offloading with FHE. The client with sensitive data generates a keypair, including evaluation key $ek$ and secret key $sk$. The server uses $ek$ to compute the encrypted data. The $sk$ is used for encrypting and decrypting ciphertexts and never leaving the client, ensuring the data confidentiality.}
    \label{fig:secure-offloading}
\end{figure}

Fully Homomorphic Encryption (FHE) enables computations to be performed directly on encrypted data.
It allows servers from untrusted cloud service providers to process sensitive information securely.
As illustrated in Figure~\ref{fig:secure-offloading}, a typical secure computation offloading scenario involves a client, who encrypts their private data, and a server, which runs the computation without access to the underlying information.
The server follows the computation protocol but lacks the secret key needed to decrypt the data.
However, computations on encrypted data are inherently intensive due to the complexity of FHE operations, making server-side computation the primary bottleneck in such systems.
After processing, the encrypted results are sent back to the client for decryption.

An FHE \textit{scheme}~\cite{bfv, fhew, chillotti_tfhe_2020, gsw, rns-ckks} defines a set of fundamental operations that enable secure computations on encrypted data, including encryption, decryption, and evaluation functions.
Among the most popular schemes are \textit{CKKS}~\cite{rns-ckks} and \textit{TFHE}~\cite{chillotti_tfhe_2020}, each suited for different types of computations.
CKKS is particularly efficient for vector operations, especially additions and multiplications~\cite{fhe_survey, ckks_regression}, making it suitable for applications like machine learning and statistical analysis.
However, CKKS faces several practical limitations.
First, it operates at the granularity of entire vectors, making operations on individual elements indirect and computationally costly.
Second, the scheme's computational pattern requires maintaining large auxiliary data throughout execution, which has led existing accelerators~\cite{craterlake, ark, bts, sharp} to incorporate substantial on-chip storage (typically 256 \texttt{MB} to 512 \texttt{MB}).

\begin{figure*}
    \centering
    \includegraphics[width=\linewidth]{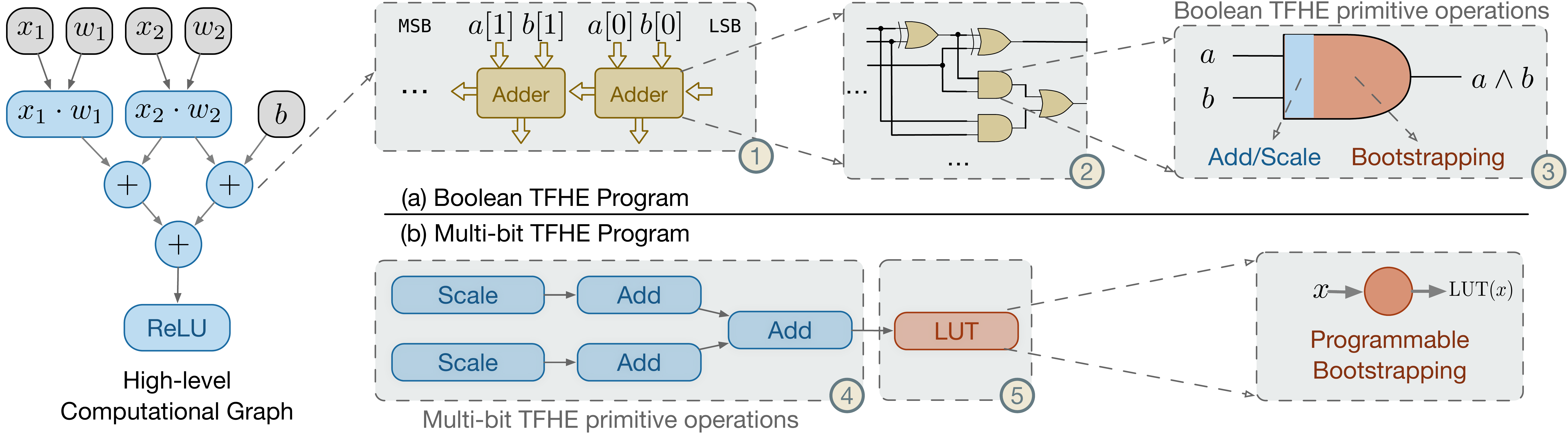}
    \caption{Breakdown of a Sample Boolean TFHE Program (a) and a Multi-bit TFHE Program (b)}
    \label{fig:multibit}
\end{figure*}

TFHE scheme ~\cite{chillotti_tfhe_2020} is often considered an alternative choice with a better balance between flexibility and performance.
TFHE supports Boolean datatype as opposed to vectors, with each bit being individually accessible after encryption, allowing the finest granularity.
With TFHE ciphertexts being more than $2000\times$ smaller than CKKS ciphertexts, TFHE accelerators ~\cite{matcha, Strix, morphling} typically require less chip area and reduced on-chip storage (generally under 50 \texttt{MB}).
Moreover, while CKKS is limited to polynomial operations (addition and multiplication) on vectors, TFHE supports logic gates that operate on Boolean data.
This enables the execution of Boolean programs that combine both polynomial and non-polynomial operations, greatly enhancing versatility compared to CKKS.
This capability has proven particularly valuable in real-world applications, from implementing convolutional neural networks ~\cite{concrete-ml-cnn} to realizing a fully homomorphic 5-stage RISC-V processor ~\cite{virtual-riscv}.

Recent works ~\cite{concrete-cnn, wop-pbs} have introduced an extension to the TFHE scheme known as \textit{multi-bit TFHE}, which allows encrypting multiple bits in a single ciphertext to form an integer.
Compared to Boolean TFHE, multi-bit TFHE often offers better performance. For example, previous work ~\cite{PyTFHE} reports an execution time of 10 minutes on image classification tasks running on a GPU. In contrast, multi-bit TFHE performs image classification in less than one second on a CPU ~\cite{concrete-ml-cnn}.

To explain this dramatic speedup, we must understand the concept of \textit{bootstrapping}.
Modern FHE schemes ~\cite{bfv, rns-ckks, chillotti_tfhe_2020} ensure security by adding noise to ciphertexts to conceal the underlying secrets.
However, in TFHE, as in other schemes, noise accumulates during computations, eventually compromising result correctness.
Bootstrapping suppresses this noise to maintain correctness but constitutes the primary performance bottleneck, accounting for over 90\% of TFHE's total runtime.

To see how the bootstrapping operation affects the runtime in a program, Figure ~\ref{fig:multibit} shows examples of how a sequence of operations (multiplications, additions, and ReLU ~\cite{first_relu}) is compiled and represented in Boolean TFHE and multi-bit TFHE respectively.

Boolean TFHE programs consist of numerous logic gates.
To compute an addition between two integers, the operation must be decomposed into binary adders operating on individual bits (step \circled{1} in Figure ~\ref{fig:multibit}(a)).
These binary adders are further broken down into various logic gates (\texttt{XNOR}, \texttt{AND}, etc.) in step \circled{2}, with each gate requiring its own bootstrapping operation (\circled{3}).
While each gate operation takes roughly 11 \texttt{ms} on a commodity processor core, image classification programs can require nearly 2.4 million gates ~\cite{PyTFHE}, thus 2.4 million bootstrapping operations, making real-time processing impractical.

Multi-bit TFHE, working directly on integers rather than binary values, enables fundamentally different program structures (Figure ~\ref{fig:multibit}(b)). 
It supports three operation types: addition, multiplication (with a constant term), and lookup tables (LUTs) for evaluating arbitrary functions. 
Additions and multiplications map directly to TFHE primitive operations as shown in \circled{4}.
This is typically seen in linear transformation layers of deep neural networks (DNN).
Non-linear operations like ReLU activation functions are encoded as LUTs (\circled{5}).

Crucially, in multi-bit TFHE, additions and multiplications can be performed without bootstrapping, making them thousands of times faster than LUT operations, which do require bootstrapping. 
Although individual bootstrapping operations in multi-bit TFHE take longer, their dramatically reduced number (only thousands for image classification versus millions) results in significantly faster overall execution.

Despite the improved efficiency of multi-bit TFHE over Boolean, a longstanding challenge remains: applications are often restricted to very short integer TFHE representations, with 4-bit being a common choice.
This limitation arises because existing hardware is significantly less efficient when handling TFHE with more than 4 bits.
For instance, evaluating a LUT on a 6-bit ciphertext is more than four times slower than on a 4-bit ciphertext on a commodity CPU.
This performance degradation stems primarily from the scaled evaluation keys (denoted as $ek$ in Figure~\ref{fig:secure-offloading}) and the auxiliary data required during evaluation, which can be roughly 4-60$\times$ larger.
The excessive bloat of the evaluation key and auxiliary data no longer fit into the L3 cache for a multi-core CPU that processes bootstrapping for several ciphertexts in parallel, making the memory bandwidth bottleneck prominent.
The resulting performance constraints impede widespread adoption of multi-bit TFHE and frequently lead to reduced efficiency when higher bit width representations are needed.

Our analysis reveals that increased bit width in TFHE necessitates scaling up internal data structures to maintain both security and correctness. 
This scaling results in data dimensions up to $30\times$ larger than those targeted by previous accelerators ~\cite{matcha, Strix, morphling}.
Previous works are either inefficient or entirely unable to accommodate this scale.
To tackle this challenge, we introduce new hardware and software techniques that, together, leverage the improved efficiency of multi-bit arithmetics.
In summary, we make the following key contributions.
\begin{itemize}
    \item We characterize multi-bit TFHE workloads, identifying a new challenge: the scaled ciphertext dimensions result in lower efficiency in previous designs and excessive memory bandwidth requirements.
    \item We propose Taurus, a hardware accelerator that scales up to 10-bit ciphertexts by shifting from many narrow PEs to fewer wider units with better FFT utilization and designing novel FFT units that efficiently handle computationally expensive functions such as external products. We reduce the memory bandwidth requirement by reusing public keys across ciphertexts.
    \item We design a compiler with multi-level operation deduplication to reuse key-switching and auxiliary data to improve memory utilization.
    \item We implement Taurus and demonstrate up to $2600\times$ and $1200\times$ speedup over commodity CPU and GPU platforms, respectively. When compared to a design employing Morphling-style systolic arrays that represents the state-of-the-art architecture, Taurus achieves up to $7\times$ higher performance.
\end{itemize}

\section{Background}

\subsection{TFHE Encryption and Internal Datatypes}

Taurus is designed to improve the FHE program execution performance of the cloud server and does not perform encryption on behalf of the clients (illustrated in Figure~\ref{fig:secure-offloading}).
However, understanding the encryption process and the internal datatypes helps readers grasp how homomorphic evaluations are performed on the server side.
Curious readers can find a more detailed discussion of the internal mechanisms of TFHE in these guides~\cite{Marc_guide, Hitchhiker_guide}.

\subsubsection{Overview of TFHE Encryption}

During initialization, the client generates a keypair.
The secret key is used locally to encrypt plaintext data of binary or integer values.
The secret key never leaves the client, preventing third parties from decrypting data.
Meanwhile, public evaluation keys are generated to enable servers to perform homomorphic operations.
In the context of TFHE, a public key comprises the \textit{bootstrapping key} (BSK) and \textit{key-switching key} (KSK), which are used by two operations described in Section~\ref{sec:tfhe-computation}.

Like any modern FHE cryptosystem, TFHE's security relies on the \emph{hardness assumption} of the learning with error (LWE) problem.
To protect ciphertexts and meet the hardness requirements, the TFHE scheme adds a small amount of noise during encryption operations to prevent data breaches by unauthorized third parties, and relies on bootstrapping operations to reduce the noise and maintain the correctness of results.
As a distinguished feature, the bootstrapping operation of TFHE is also \emph{programmable}, and thus is referred to as programmable bootstrapping (PBS).
The \emph{programmability} of PBS refers to its capability to evaluate a univariate function, often referred to as a lookup table (LUT) in this context.
One can think of PBS as an evaluation function that also reduces the noise to an acceptable level.

\subsubsection{Internal Datatypes in TFHE}

\label{sec:internal-datatypes}

In TFHE, torus elements are central to the encryption scheme, and are the fundamental building block for the internal datatypes.
In the context of TFHE, a \textit{torus} $\mathbb{T}$ refers to a 1-D torus, or circle.
Conceptually, torus elements are real values in the range $[0,1)$.
In practice, these are represented as \textit{discretized} torus elements - specifically, the set of $2^{w}$ $w$-bit fixed-point fractions in $[0,1)$, with $w$ usually being 32 or 64.

Since PBS is the most computationally extensive operation that dominates the overall runtime, Taurus's design choices are tightly related to improving the efficiency of PBS operations, which internally relies on three datatypes, referred to as \textit{LWE}, \textit{GLWE}, and \textit{GGSW}.

The \textit{LWE} ciphertexts ~\cite{chillotti_tfhe_2020} are used by clients to encrypt messages.
This ciphertext has the smallest size along the three types and is parameterized by its dimension $n$, typically ranging from 500 to 1000.
Each LWE ciphertext has $n+1$ torus elements: $n$ elements called the LWE \textit{body} plus an extra one called the LWE \textit{mask}.

The Generalized-LWE (GLWE) ~\cite{chillotti_tfhe_2020} ciphertexts are used to encode the LUT for PBS and store the intermediate results during bootstrapping.
GLWE ciphertexts generalize from LWE ciphertexts by replacing each torus scalar with a polynomial, where each polynomial has a degree that is a power of two, denoted as $N$.
There are $k+1$ polynomials in each GLWE ciphertext, where $k$ is called the \textit{GLWE dimension}.

Generalized Gentry-Sahai-Waters (GGSW) ciphertexts ~\cite{gsw} are used in the bootstrapping key (BSK).
It is a matrix of polynomials.
Each BSK consists of $n$ GGSW ciphertexts.
The main purpose of GGSW is to enable multiplication between torus polynomials (in GLWE ciphertexts) and integer polynomials (in GGSW ciphertexts), which is required for bootstrapping.
This multiplication operation is referred to as the \textit{external product}, denoted as $\text{GGSW} \boxdot \text{GLWE} \rightarrow \text{GLWE}$.
Specifically, each GGSW ciphertext is a matrix of $(1+k)^2 \times d$ polynomials, where $d$ stands for \textit{decomposition depth}.

\subsection{TFHE Computation}

\label{sec:tfhe-computation}

The computations in a TFHE program can be seen as a combination of linear operations and PBS operations.
Linear operations are fast, involving simple elementwise additions or multiplications.
This section explains how the ciphertexts introduced in Section ~\ref{sec:internal-datatypes} interact in a PBS operation.

\begin{figure}[t]
    \centering
    \includegraphics[width=\linewidth]{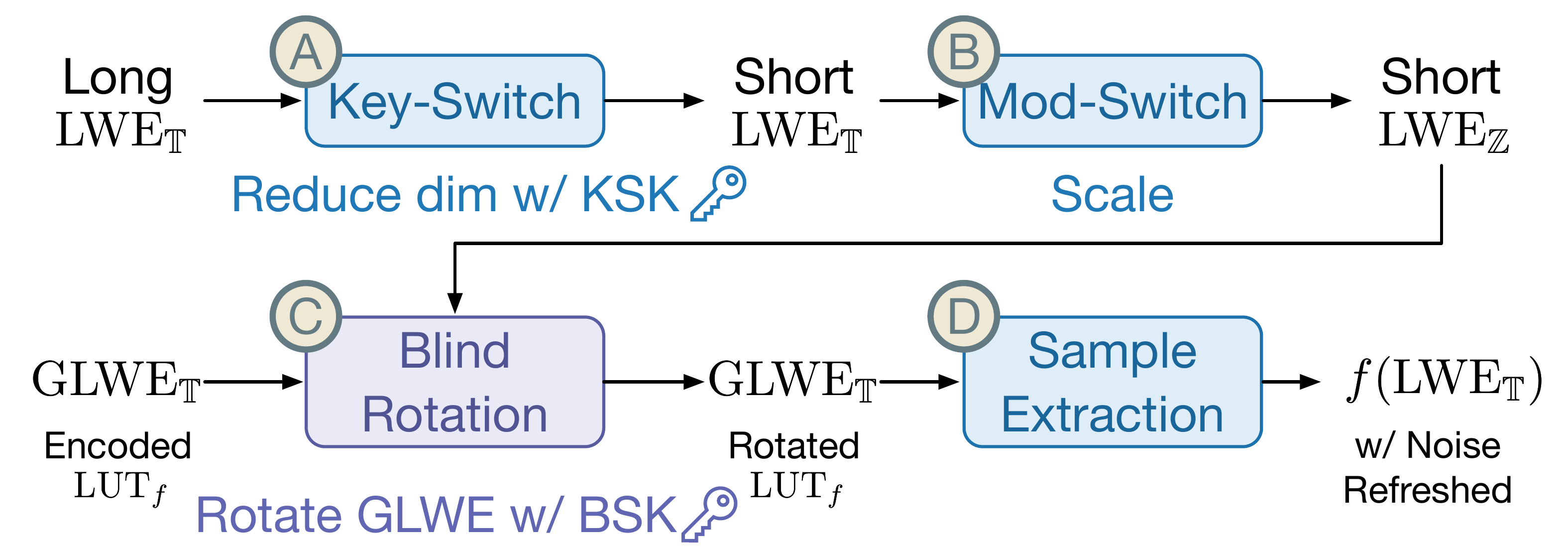}
    \caption{Outline of TFHE Programmable Bootstrapping}
    \label{fig:pbs}
\end{figure}

Figure~\ref{fig:pbs} illustrates the major steps in a PBS operation.
From the input and output perspective, the input of PBS is an LWE ciphertext.
The output is an LWE ciphertext with the same dimension, reduced noise, and with function $f$ evaluated on it.
The PBS process consists of four main steps.

The \textit{key-switching} operation (shown as \circled{A}) reduces the dimension of an input LWE ciphertext (denoted as \textit{long} to \textit{short} in Figure~\ref{fig:pbs}) by using the key-switching key (KSK).
This dimension reduction is desirable because the reduced dimension results in fewer iterations in blind rotation (\circled{C}), and thus reducing the total computations (i.e. from 30,000 to 1000).
Computationally, the core of key-switching is to iteratively multiply the KSK with LWE ciphertexts.
Key-switching is the second most time-consuming operation, usually taking less than 10\% of the total runtime.

The \textit{modulus-switching} (Mod-Switch, shown as \circled{B}) performs type conversion for LWE ciphertexts.
Specifically, it scales and rounds the LWE ciphertext values from torus elements in $\mathbb{T}$ to integers ($\mathbb{Z}$), which is required for the blind rotation step.
Mod-Switch is generally a fast operation, taking less than 1\% of the total runtime.

\begin{figure}[t]
    \centering
    \includegraphics[width=\linewidth]{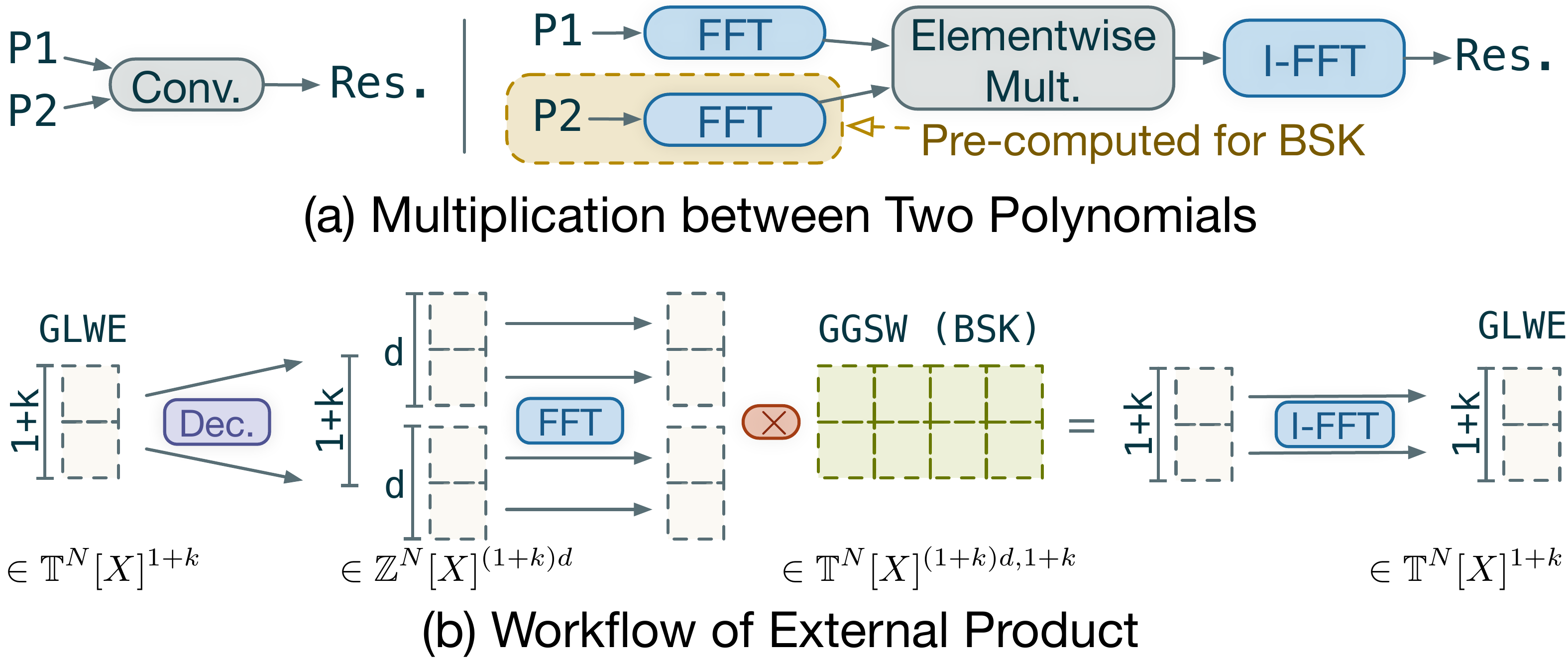}
    \caption{Multiplication between Polynomials and its Usage in External Product}
    \label{fig:ext-product}
\end{figure}

The \textit{blind rotation} (\circled{C} in Figure~\ref{fig:pbs}) takes a GLWE ciphertext that encodes a lookup table (LUT), and transforms it based on the mod-switched LWE ciphertext.
It is the most time-consuming part because it iteratively performs multiplication between large polynomials.
Figure~\ref{fig:ext-product} (a) presents two methods for polynomial multiplication in general.
Naively, it requires discrete convolutions (denoted as \texttt{conv.}) between the coefficients.
In practice, FFT and inverse FFT are used to speedup computation by reducing the overall computational complexity.

The core of blind rotation is the iterative \textit{external products} between GLWE (which encodes a LUT) and GGSW ciphertexts.
The workflow of external product is shown in Figure~\ref{fig:ext-product} (b).
Each external product is essentially a vector-matrix multiplication between the BSK and decomposed GLWE ciphertexts, where each element is a large polynomial with degree $N$.

Finally, the \textit{sample extraction} step (shown as \circled{D} in Figure~\ref{fig:pbs}) extracts an LWE ciphertext from these constant terms, restoring it to the same dimension as the original input to the PBS.
This is a fast function that typically takes less than 1\% of the runtime.

Note that there are two possible execution orders in PBS.
Besides the key-switching-first order of operations shown in Figure~\ref{fig:pbs} (\circled{A} $\rightarrow$ \circled{B} $\rightarrow$ \circled{C} $\rightarrow$ \circled{D})~\cite{ks_first}, an alternative widely adopted order is blind rotation-first (relative to key-switching, \circled{B} $\rightarrow$ \circled{C} $\rightarrow$ \circled{D} $\rightarrow$ \circled{A})~\cite{chillotti_tfhe_2020}.
While the order does not change the total amount of computation required for a single PBS operation, the key-switching first approach offers advantages in multi-bit TFHE programs.
Specifically, when multiple PBS operations are needed in sequence, the key-switching-first order creates opportunities for data reuse and deduplication across operations.
For this reason, we adopt the key-switching-first order throughout this work.

\section{Motivation}
\label{sec:motiv}
In this section, we examine the key challenges and opportunities of executing multi-bit TFHE programs on hardware accelerators.
Throughout the paper, we present several key observations and insights into various hardware design choices for multi-bit TFHE.
These insights offer guidelines for shaping the Taurus architecture and the future development of TFHE accelerators.

\subsection{Efficiency of Multi-Bit TFHE Programs}

\subsubsection{Programmability of Bootstrapping}

TFHE was initially designed to support the encoding and evaluation of Boolean variables where each ciphertext encodes a single Boolean value (\texttt{true} or \texttt{false}). 
Since TFHE operates on a discretized torus ranging from 0 to 1, an additional abstraction layer (gates) divides this torus into two halves, with one half representing \texttt{true} and the other \texttt{false}.

This abstraction layer prevents Boolean TFHE from leveraging a key feature of the TFHE scheme: the programmability of bootstrapping.
Instead of utilizing PBS's ability to encode constant values in lookup tables (LUTs) at no additional computational cost compared to simply refreshing the noise, Boolean TFHE embeds these constants into its computational graph.
Recall that execution of Boolean TFHE programs essentially involves evaluating a computational graph composed of different types of homomorphic gates (i.e. \texttt{NAND}).
Internally, each homomorphic gate performs two key steps: first combining two encrypted bits using a linear operation (depending on the type of gate), then applying a bootstrapping operation to manage noise growth.
The bootstrapping operation remains mandatory for every gate throughout the program's execution.
As a result, the high-level linearity in a Boolean TFHE program (such as multiply-accumulate (MAC) operations) cannot be transferred to low-level bootstrapping-free TFHE primitive linear operations.

\paraobserv{Observation 1: Boolean TFHE's practice of encoding constants within its computational graph, rather than utilizing the zero-overhead programmability of PBS through lookup tables, explains its lower execution efficiency compared to multi-bit TFHE programs.}

Multi-bit TFHE, by contrast, takes full advantage of the programmability by encoding constant information within LUTs (as opposed to encoding into the computational graph) while mapping high-level linear operations (such as MACs) directly to the linear primitives that do not depend on bootstrapping.

\begin{figure}[t]
    \centering
    \includegraphics[width=\linewidth]{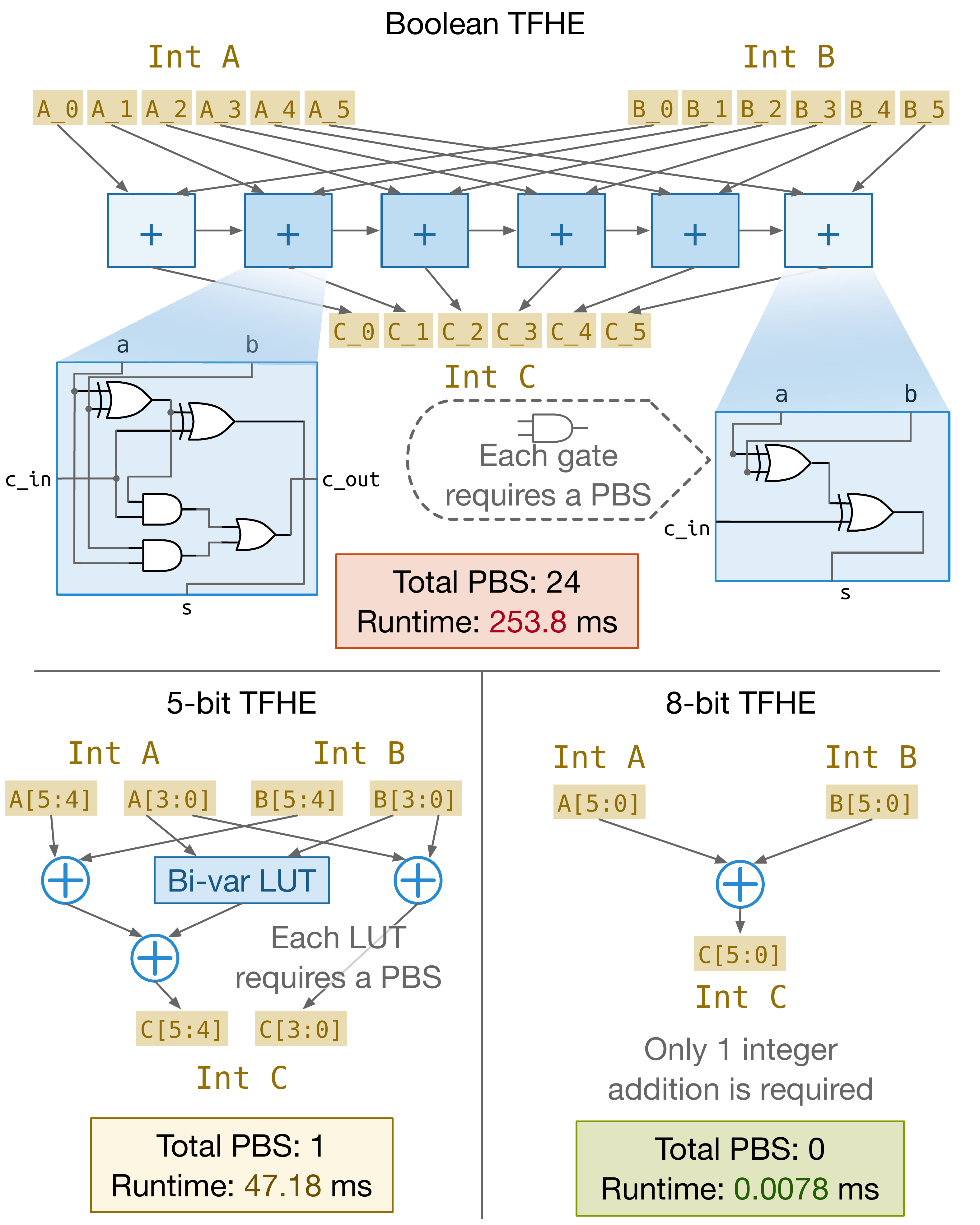}
    \caption{Performing a 6-bit integer addition using various TFHE representations.}
    \label{fig:addition}
\end{figure}

\subsubsection{Wider Numeric Representations}

Besides preserving the linearity, the removal of the abstraction layer creates another chance to optimize performance through wider numeric representations, which is not applicable to Boolean TFHE.

While the complexity of each individual PBS operation grows quadratically with the numeric width, the broader perspective of the entire program reveals a crucial insight: wider representations often require fewer total PBS operations and, in turn, improve the overall runtime.

To demonstrate how wider numeric representations can improve execution efficiency, Figure~\ref{fig:addition} presents a comparative analysis of adding two 6-bit integers using Boolean, 5-bit, and 8-bit TFHE representations, respectively.
Let's examine each representation's performance for a 6-bit integer addition:

Using Boolean TFHE (Figure~\ref{fig:addition}-top), we implement a ripple carry adder, which provides the most gate-efficient design for TFHE programs.\footnote{While other adder architectures like carry-lookahead often offer better performance in traditional hardware, they require more gates, making them less efficient for TFHE where each gate requires expensive bootstrapping operations.}
Even though each individual Boolean logic gate executes in just 11~\texttt{ms} on a commodity CPU core, the total number of gates needed for the complete addition results in an execution time of 253~\texttt{ms}.\footnote{Unless stated otherwise, all measurements in this section are performed using the TFHE-rs library (commit \texttt{4c9b081}) on AMD EPYC 7R13.}
Moving to 5-bit TFHE (Figure~\ref{fig:addition}-bottom left), the approach requires splitting each 6-bit integer into two segments according to the radix.\footnote{While the Chinese remainder theorem (CRT) could theoretically avoid carry propagation between ciphertexts, its limited support for different operation types has prevented widespread adoption in complex real-world applications.}
Though adding individual segments proceeds quickly, managing the carry bit between segments requires implementing a bivariate LUT.\footnote{Bivariate LUT is not a TFHE-native operation. It is commonly implemented by first combining two ciphertexts linearly, then applying a standard univariate LUT to the result.}
This LUT operation necessitates one PBS operation and becomes the performance bottleneck, leading to a total runtime of 47~\texttt{ms}.

The 8-bit TFHE approach (Figure~\ref{fig:addition}-bottom right) demonstrates the advantage of wider representations: a single 8-bit ciphertext can fully accommodate a 6-bit integer.
This eliminates the need for carry propagation between ciphertexts and achieves remarkably faster execution at just 0.008~\texttt{ms}.

\paraobserv{Observation 2: Although individual operations (i.e. LUTs or bootstrappings) with wider multi-bit TFHE representations incur higher costs than those with narrower representations, the reduction in the total number of required operations often leads to significantly better overall program performance.}

\subsection{Challenges of Executing Multi-bit TFHE Programs}

Each TFHE program is associated with some parameter sets.
The choice of parameter set jointly impacts the robustness of operations, the computational cost, and the supported message width while maintaining the desired security level.

\begin{figure}
    \centering
    \includegraphics[width=\linewidth]{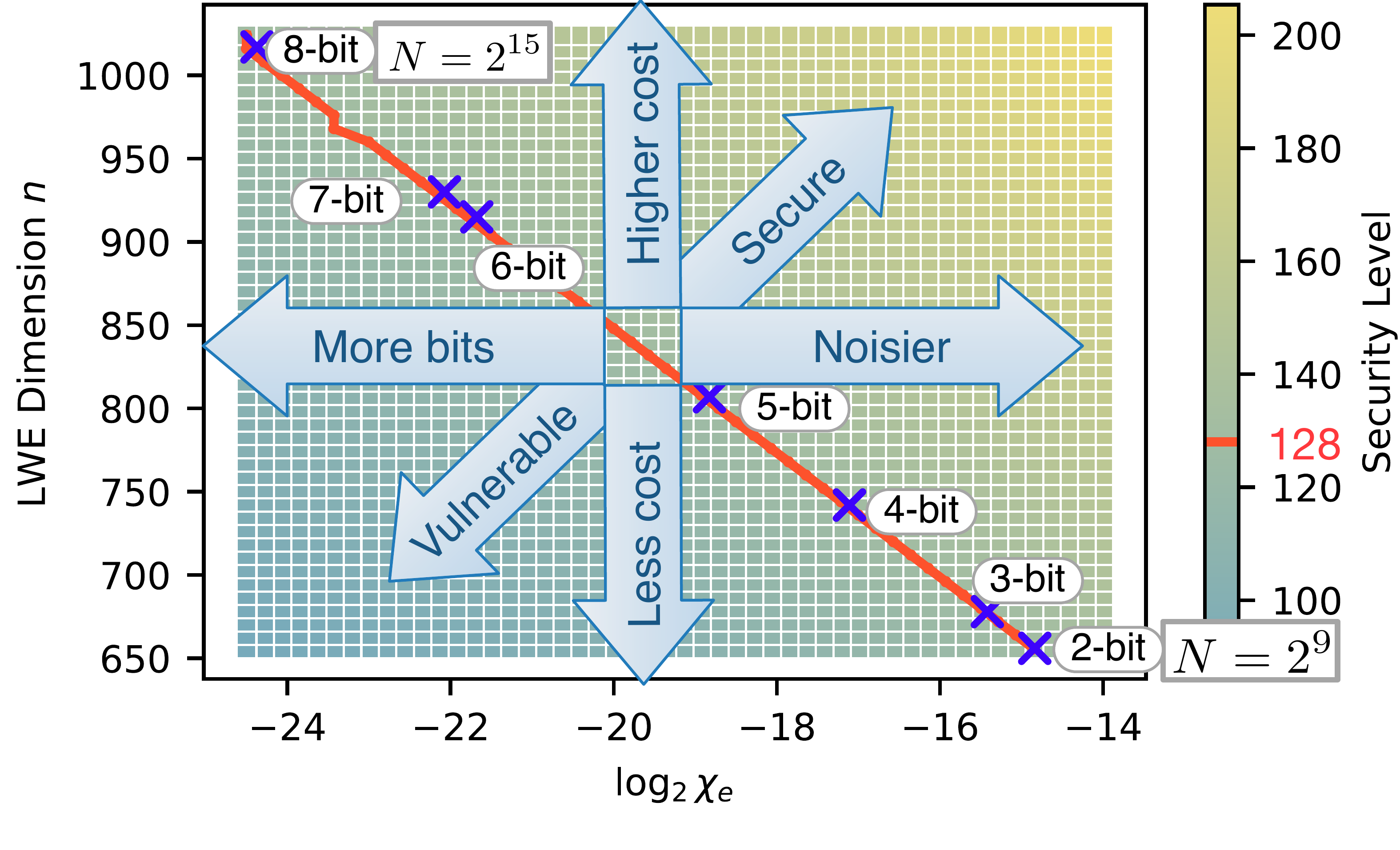}
    \caption{Interplay between parameters. Supporting a larger width requires a larger dimension to maintain the desired security level (128-bit). Arrows indicate the orthogonal goals of security, robustness, and performance.}
    \label{fig:grid_search}
\end{figure}

\textbf{The challenge of wide numeric representation is associated with the growth of parameter sets.}
While Boolean TFHE operates with a fixed, stable parameter set across all programs, multi-bit TFHE faces a more complex scenario.
As the number of bits encoded in each ciphertext becomes variable, the parameter search space expands significantly.

Using the Lattice Estimator~\cite{lwe_estimator} to evaluate potential cryptographic attacks, we analyzed the relationship between security levels and parameter choices, as shown in Figure~\ref{fig:grid_search}.
The y-axis denotes the LWE dimension $n$, and x-axis denotes $\chi_e$, the standard deviation of the noise distribution.
The parameter combinations that achieve 128-bit security are marked, forming a red line on the figure.\footnote{$n$-bit security indicates that the best known attacks require at least $2^n$ operations to break the instance.}
Parameter $n$ must increase to support wider representations while maintaining the desired security level, which directly increases the computational complexity of bootstrapping.\footnote{$\chi_e$ must decrease as more bits are added to maintain decryption correctness.}
Figure~\ref{fig:grid_search} also highlights parameter sets for different bit widths,\footnote{These parameters ensure a failure probability of $p_{\text{error}} < 2^{-40}$, small enough to be negligible in practice.} demonstrating the interdependence between LWE and GLWE parameters.
For example, the doubled LWE dimension $n$ corresponds to a $64\times$ growth of the GLWE polynomial degree $N$.
Any architecture for multi-bit TFHE must therefore handle these expanded parameter sets efficiently while maintaining security guarantees.

\begin{figure}
    \centering
    \includegraphics[width=0.95\linewidth]{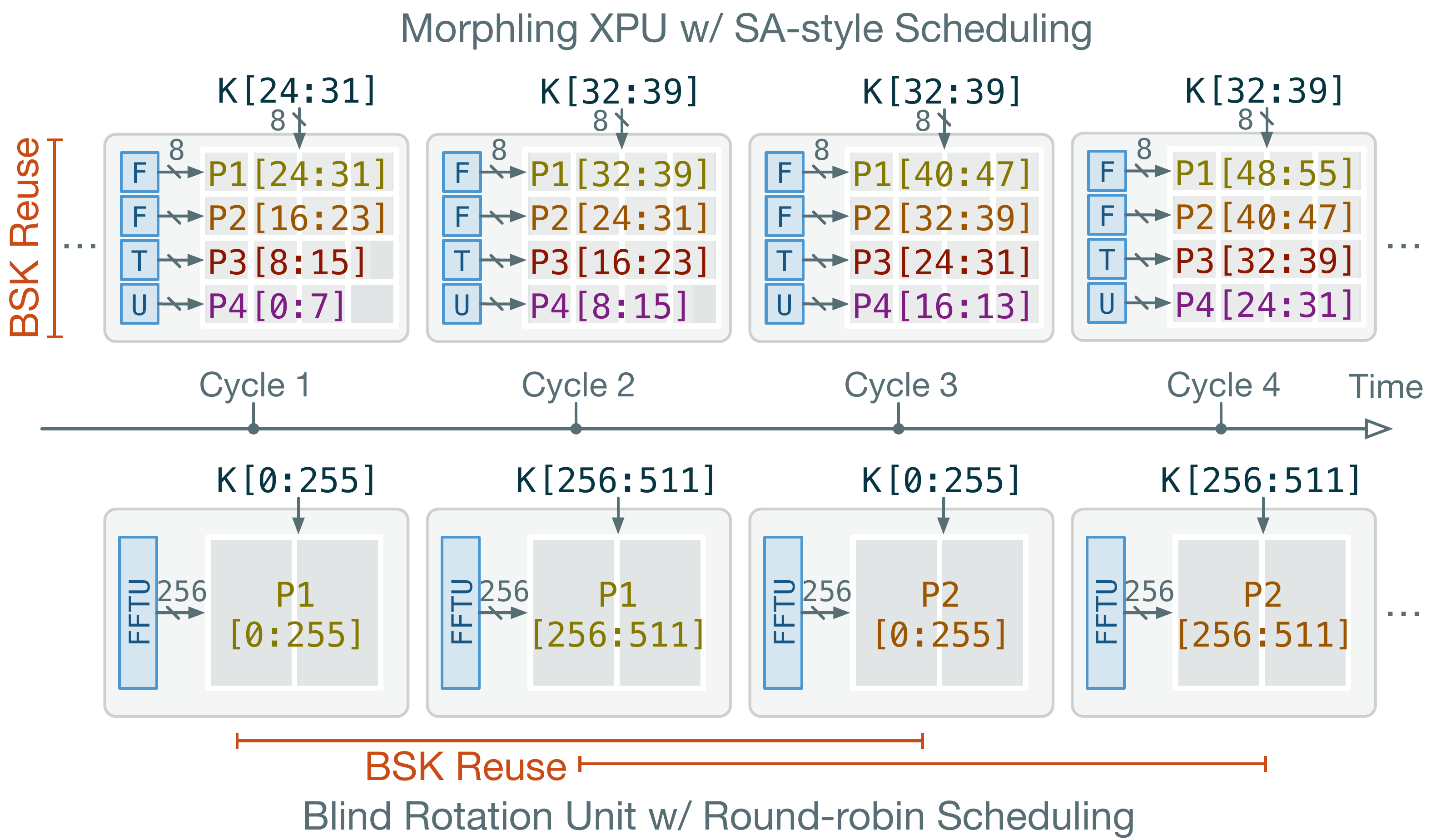}
    \caption{Comparison of polynomial multiplication execution patterns and key reuse strategies between Morphling's External Product Unit (XPU) and Taurus's Blind Rotation Units (BRU).}
    \label{fig:XPU-vs-BRU}
\end{figure}

\textbf{The challenge of simply scaling up existing TFHE accelerators for wider bit-width representation support.}
State-of-the-art TFHE accelerators use systolic array-like design to perform external product, the most time-consuming operations in bootstrapping.
Figure~\ref{fig:XPU-vs-BRU}-top shows how Morphling, the state-of-the-art accelerator, performs FFT and multiplications on polynomial chunks.
Each FFT unit (denoted as \texttt{FFTU} in the figure) produces 8 coefficients per cycle.
Four rows, each with an FFTU, process four independent polynomials (denoted as \texttt{P1} to \texttt{P4}).
The processing elements (PE) of each row perform elementwise multiplications between coefficients from FFT units on the left, and chunks of BSK from the top.
The results accumulate inside each PE, featuring an output-stationary design.
However, the architecture faces fundamental challenges when scaling up to adapt programs with wider numeric representations.

\emph{Horizontally scaling systolic arrays does not improve data reuse in multi-bit TFHE program execution and reduces utilization instead.}
Morphling employs 4 PEs in a row.
Among each row in the systolic array (denoted as \texttt{SA} in Figure~\ref{fig:XPU-vs-BRU}), outputs from FFTU are broadcasted and reused across all PEs.
Programs with GLWE dimension $k$ can utilize up to $k+1$ PEs, which enables Morphling to efficiently handle up to $k=3$ with 4 PEs in a row.
However, while $k=2$ or $k=3$ is common in Boolean TFHE and low-width TFHE (i.e., up to 3 bits), our observation shows that wider-width TFHE typically sets $k=1$.
This is mainly because the complexity of PBS grows quadratically with $k+1$,
making large $k$ less favorable.
This mismatch (more PEs in a row than needed) leads to significant inefficiency: when processing wider-width TFHE operations, 50\% of the PE array, which consumes the majority of chip area, remains idle and negatively impacts the throughput per unit area.
The reuse opportunity that is common for low-width TFHE becomes less significant for wider-width TFHE.

\emph{Vertically scaling systolic arrays requires excessive storage for accumulator data.}
Systolic arrays reuse on-chip BSK chunks (denoted as \texttt{K} in Figure~\ref{fig:XPU-vs-BRU}) by passing them down along each column.
Since each BSK coefficient is used exactly once per PBS operation, vertical scaling primarily benefits parallel processing of multiple ciphertexts rather than improving single-ciphertext throughput.
However, this approach introduces significant storage overhead because each ciphertext requires its own accumulation data for blind rotation, as increasing the number of rows proportionally increases the required accumulator storage.
Additionally, vertical scaling necessitates duplicating FFTU units across rows, as each row requires its own dedicated FFTU.
This FFTU duplication is less area-efficient compared to increasing the throughput of individual FFTU units.

\paraobserv{Observation 3: Assuming an output-stationary, systolic-array design where GLWE intermediate ciphertexts serve as accumulators, fewer high-throughput or wide PEs are more efficient than numerous narrow PEs, primarily due to reduced intermediate storage requirements and more efficient FFT unit utilization.}

\emph{Individual PE scaling is also limited by memory bandwidth constraints due to the BSK reuse strategy of systolic arrays.}
Although making each PE more powerful might seem like a solution, our analysis reveals a critical bottleneck: BSK is reused between PEs.
With no opportunity for BSK reuse within an individual PE, increasing throughput requires proportionally more BSK elements as inputs.
In fact, even a modest 2$\times$ scaling of the PEs would saturate Morphling's available memory bandwidth.

These fundamental scaling limitations of existing architectures---horizontal, vertical, and per-PE, motivated a different design principle that achieves higher-throughput polynomial multiplication (PolyMult) operations while supporting larger polynomial degrees that are necessary for high-width multi-bit TFHE workflows.

\textbf{Design principle of Taurus.}
Instead of having multiple polynomials performing low-throughput PolyMult operations independently and concurrently, Taurus favors fewer polynomials performing higher-throughput PolyMult operations, which enables the design of FFT units with better throughput per unit area.
To facilitate BSK reuse and reduce memory bandwidth requirements, Taurus employs round-robin scheduling to reuse the same key chunks across multiple polynomials.
As shown in Figure~\ref{fig:XPU-vs-BRU}-bottom, Taurus loads only one polynomial into its FFT pipeline per core in each cycle, maximizing the throughput of individual FFT units.
Additionally, the round-robin approach allows multiple independent polynomials to take turns performing FFT and MAC operations, enabling efficient BSK reuse across ciphertexts.
Taurus's architecture scales efficiently in throughput, number of cores, and supported TFHE program bit width (up to 10 bits), facilitating more flexible parameter choices and better performance for multi-bit TFHE programs.

\section{Taurus Architecture}
\label{sec:arch}

\subsection{Top-level Organization}

\begin{figure}
    \centering
    \includegraphics[width=\linewidth]{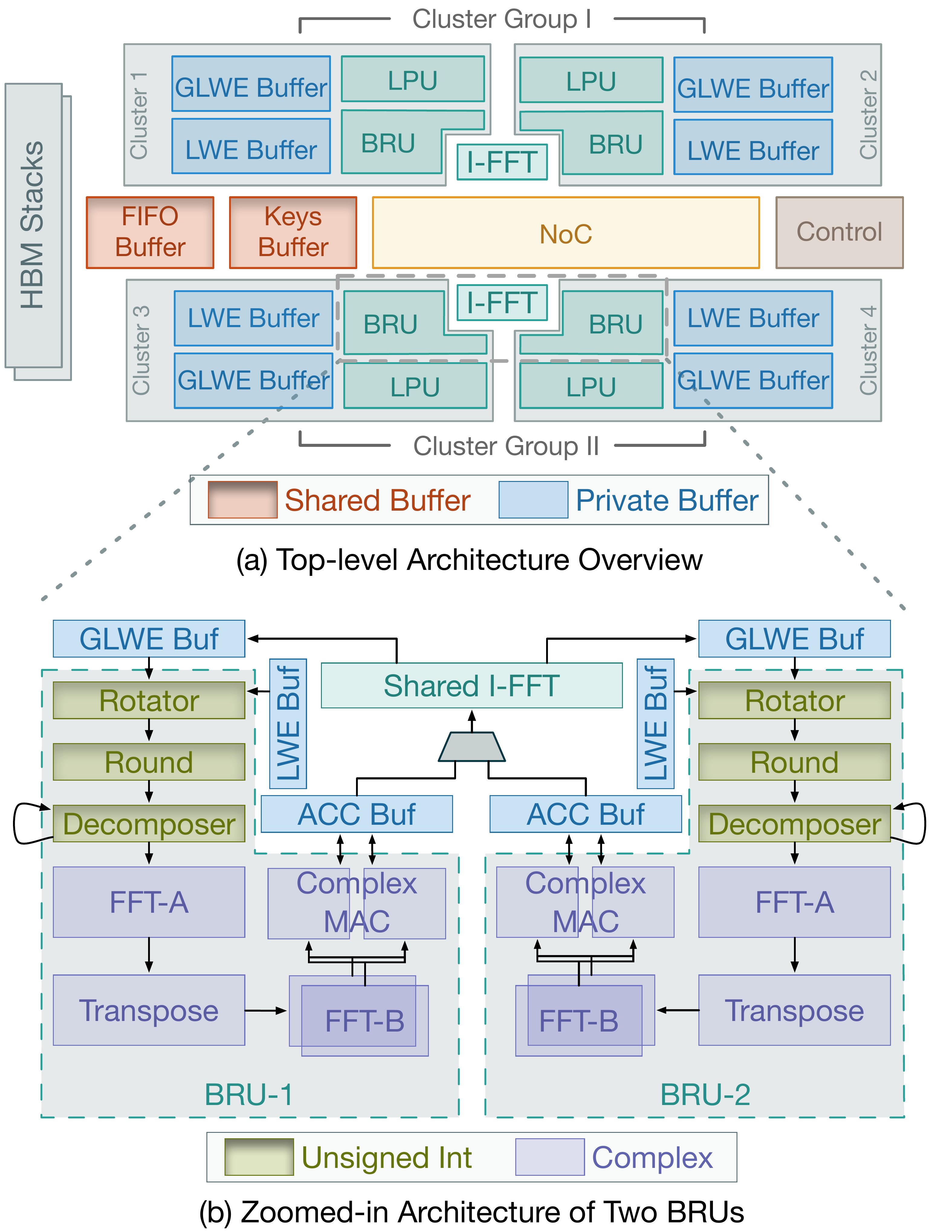}
    \caption{Logical Organization of Taurus Architecture and Breakdown of Compute Clusters (a) and Zoomed-in Organization of Two Blind-rotation Units (b)}
    \label{fig:arch-wide}
\end{figure}

Taurus's architecture, shown in Figure~\ref{fig:arch-wide}(a), is built around four vector-core-like compute clusters.
Each cluster incorporates two specialized functional units: a Blind-rotation Unit (BRU) and an LWE Processing Unit (LPU).

\textbf{Blind-rotation Unit (BRU)} implements a deeply pipelined design optimized for processing streams of complex numbers.
As shown in Figure~\ref{fig:arch-wide}(b), two BRUs in each cluster share a single inverse FFT (IFFT) unit.
Each BRU achieves a throughput of 512 BSK multiplications per cycle for blind rotation operations.

Through extensive testing with both TFHE-rs library parameters and Concrete Optimizer-generated parameters\footnote{All parameter sets maintain 128-bit security level.}, we determined that 48-bit fixed-point numbers optimally represent the real and imaginary components of complex numbers.

\paraobserv{Observation 4: 48-bit fixed-point data type ensures correctness across all tested parameter sets while maintaining compatibility with both TFHE-rs and Concrete Optimizer.} 

\textbf{LWE Processing Unit (LPU)} specializes in operations on LWE ciphertexts, including key-switching, modulus-switching, addition, and plaintext multiplication.
The LPU operates with a 64-bit width to match the torus modulus of $2^{64}$.
It includes a vector addition and multiplication unit that performs homomorphic addition and plaintext multiplication through element-wise operations.
The most computationally expensive task in the LPU is key-switching, which involves multiple levels of decomposition of a long-LWE ciphertext.
Since the length of a short-LWE ciphertext is not necessarily a power of two, the LPU is designed with four parallel lanes, each capable of processing 64 elements instead of a single wide lane adopted in the BRU.\footnote{Our testing confirms that four lanes are enough to complete key-switching and the associated linear operations before blind rotation finishes across all tested parameter sets.}
Each lane handles one decomposed LWE scalar.

\textbf{Memory Subsystem} of Taurus employs a hierarchical on-chip memory consisting of global buffers shared by all clusters and per-cluster private buffers.
The sequential access patterns enable effective DRAM latency hiding using a modest 16~\texttt{KB} read/store queue.
The global buffers store portions of BSK and KSK that are shared across all clusters and distributed via the NoC.
The private buffers consist of a GLWE accumulator buffer (complex numbers) accessed exclusively by the BRU, and GLWE and LWE buffers (unsigned integers) shared between the BRU and LPU.

\subsection{Frontend and Control}

\textbf{Data Flow and Operation Scheduling}

\begin{figure}
    \centering
    \includegraphics[width=\linewidth]{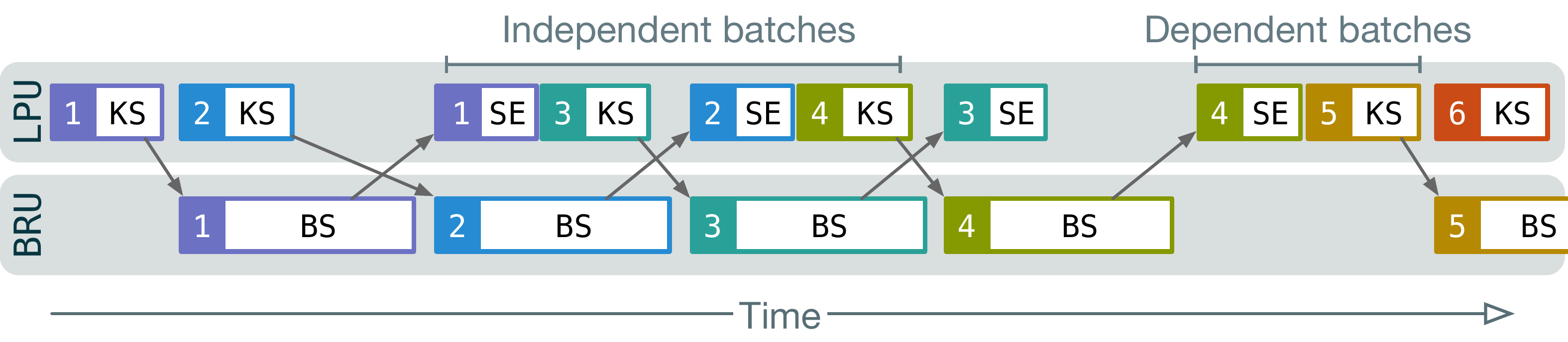}
    \caption{Scheduling of operations across BRU and LPU modules for independent and dependent ciphertext batches.}
    \label{fig:FU-scheduling}
\end{figure}

Taurus schedules operations at the batch granularity, with each batch containing up to 48 ciphertexts across cores (12 ciphertexts per core).
Figure~\ref{fig:FU-scheduling} shows how operations from 6 batches are scheduled to the LPU and BRU, respectively.\footnote{For readability, modulus-switching and linear operations are not shown. These are handled by the LPU. Execution times are not drawn to scale.}
Each block indicates the batch number and operation type.
Our proposed compiler (Section~\ref{sec:compiler}) groups ciphertexts into batches and schedules them based on data dependencies.
For independent batches such as batches 1-4, the compiler overlaps bootstrapping (BS) operations on the BRU with other operations like key-switching (KS) and sample extraction (SE) on the LPU.
However, when consecutive batches have data dependencies, such as batches 4 and 5, the BRU must wait for the previous key-switching operation to complete.

\begin{figure*}
    \centering
    \includegraphics[width=\linewidth]{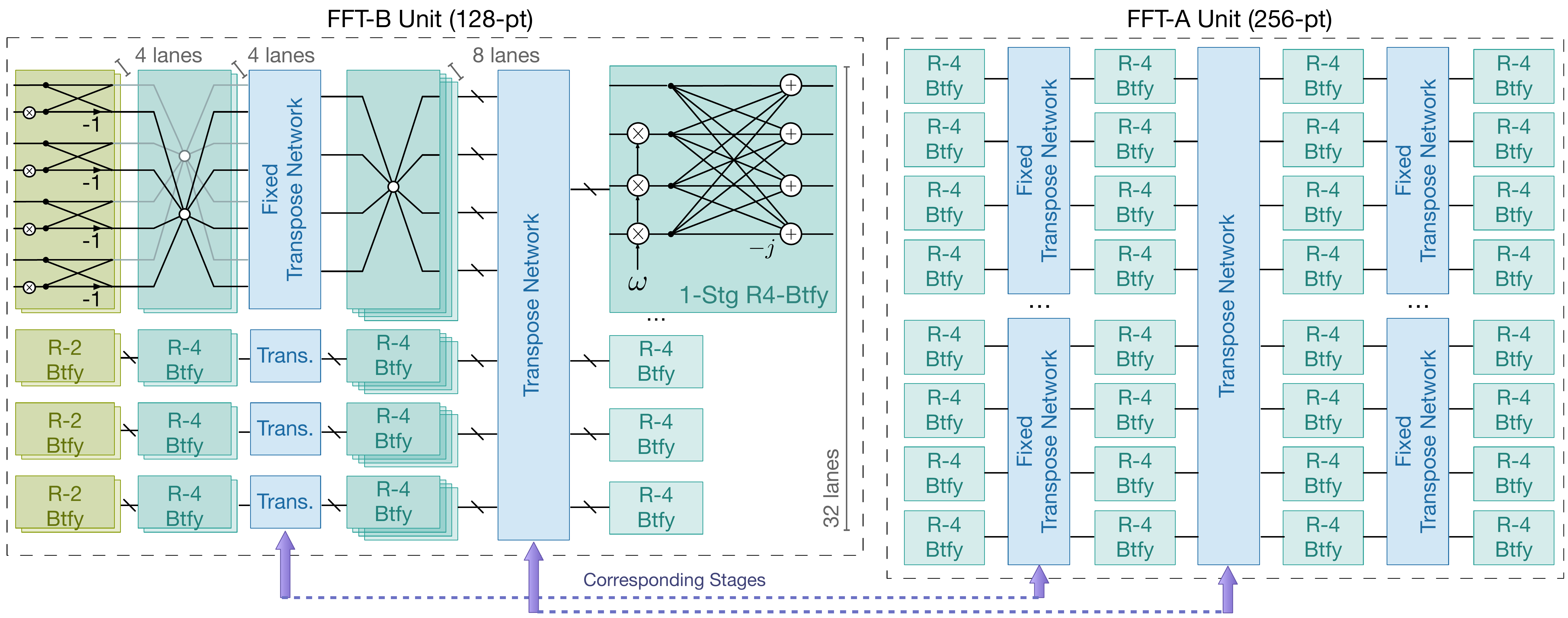}
    \caption{TFHE-tailored FFT unit design. Two types of FFT clusters are used to match the decomposition requirement of multi-bit TFHE parameters. The 256-pt cluster has a symmetric design, whereas the 128-pt cluster is asymmetric. The correspondence between the stages is indicated.}
    \label{fig:fft-wide}
\end{figure*}

\textbf{Synchronization Strategy}

Taurus adopts an aggressive full synchronization strategy, where all clusters are synchronized to perform blind rotation and key-switching in the same iteration.
This approach maximizes key reuse and minimizes memory bandwidth requirements.
However, it introduces a limitation: new bootstrapping operations must wait until all current operations complete across all clusters, even when some clusters have finished their linear operations and are ready to proceed.

To explore the impact of synchronization strategies, we implemented a grouped synchronization approach that organizes clusters into up to two groups, allowing different groups to execute operations independently while maintaining intra-group synchronization.
Evaluation across various workloads shows that grouped synchronization provides minimal benefits: median speedup of only 0.07\% (maximum 3.53\%) while nearly doubling peak memory bandwidth from 734~\texttt{GB/s} to 1449~\texttt{GB/s}.

\paraobserv{Observation 5: 
While grouped synchronization offers a small reduction in runtime, it significantly increases memory bandwidth demands.
Additional flexibility does not translate into substantial performance improvements and comes with a high bandwidth cost.}

\subsection{TFHE-Tailored Functional Unit for (I)FFT}

A key challenge in multi-bit TFHE is handling the significantly larger polynomial degrees required to support wide numeric representations.
Taurus supports polynomials up to degree $2^{16}$, sufficient for ciphertexts that encrypt up to 10 bits with 128-bit security.

Our design employs double-real FFT~\cite{double-real2}, which efficiently processes a $2^{16}$-degree polynomial using only a ${2^{15}}$-point complex vector.
However, the $2^{15}$-point sequence presents a unique challenge.
Prior work~\cite{f1, craterlake, sharp} relies on decomposition that applies to perfect-square vector lengths (e.g., CraterLake's~\cite{craterlake} choice of $256 \times 256$, which is a perfect square).
However, a $2^{15}$-point sequence cannot be evenly divided into equal-sized parts (like $\sqrt{N} \times \sqrt{N}$) and mapped to homogeneous functional units.
This constraint led us to develop a novel heterogeneous functional unit design, which is a key distinguishing feature of Taurus compared to previous FFT/NTT implementations.
Each FFT cluster employs two types of functional units: FFT-A processes 256-point sequences and FFT-B processes 128-point sequences, interconnected by a transpose unit.

As shown in Figure~\ref{fig:fft-wide}, the FFT-A unit uses a symmetric design with $\sqrt{256}$ lanes and 16 elements per lane.
The FFT-B unit features an asymmetric design based on decomposing into four 32-point sequences, which are further decomposed into four 8-point sequences.
Both units use a mixture of radix-2 and radix-4 butterfly units, with radix-4 saving 25\% complex multiplications compared to radix-2.
Every stage can be optionally bypassed to support various sequence lengths, with the initial radix-2 stage providing flexibility for lengths like $2^{14}$.

Compared to the 8-parallel R2MDC design used by the state-of-the-art TFHE accelerator~\cite{morphling}, our heterogeneous FFT cluster uses $1.38\times$ the area while achieving $32\times$ better throughput.

\subsection{Transpose Unit}

\begin{figure}
    \centering
    \includegraphics[width=\linewidth]{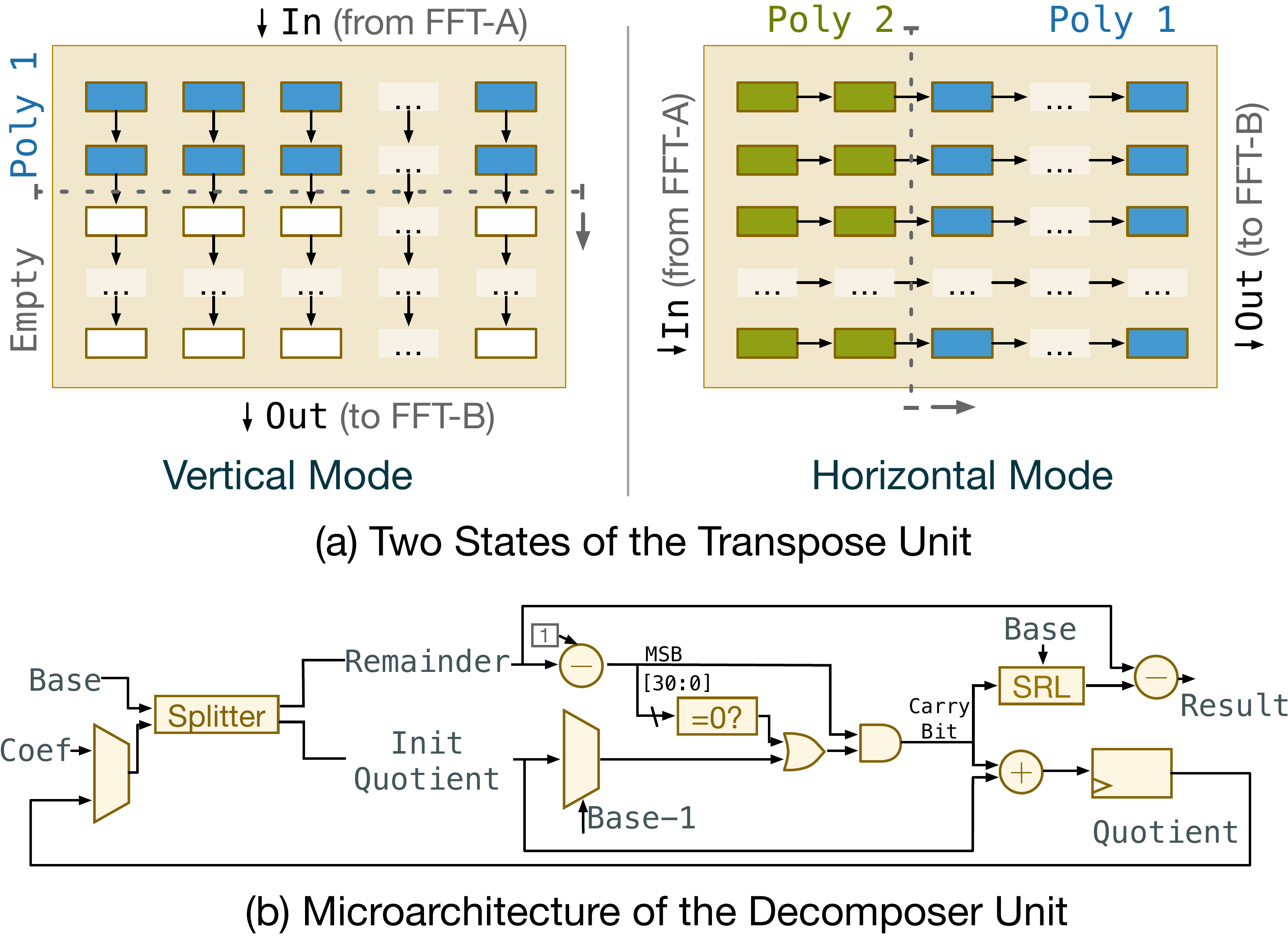}
    \caption{States of the Transpose Unit (a) and microarchitecture of the Decomposer Unit (b)}
    \label{fig:transpose}
\end{figure}

Transpose operations occur in two contexts: within individual FFT units and between FFT-A and FFT-B units. For the latter case, the FFT-B unit must wait up to 128 cycles for FFT-A to process all data within a polynomial, creating throughput challenges. We designed a "Shutter Transpose" unit that streams data either horizontally or vertically using internal counters to track polynomial boundaries, similar to camera shutter curtains.
As shown in Figure~\ref{fig:transpose}(a), the unit alternates between vertical streaming for incoming polynomials and horizontal streaming for outgoing data to maintain constant throughput.

\subsection{Decomposer Unit}

The decomposition process converts each element in a torus polynomial into a vector of integers by representing the element in a power-of-two base $B$ across decomposition depth $d$.
As shown in Figure~\ref{fig:transpose}(b), our hardware implementation consists of two components: an initial scaling unit that may introduce stalls for decomposition depths greater than 1, and a continuous digit extraction unit that outputs one integer per cycle with built-in rounding logic to maintain the required FFT cluster throughput.

\section{Compiler}
\label{sec:compiler}

\begin{figure}
   \centering
   \includegraphics[width=\linewidth]{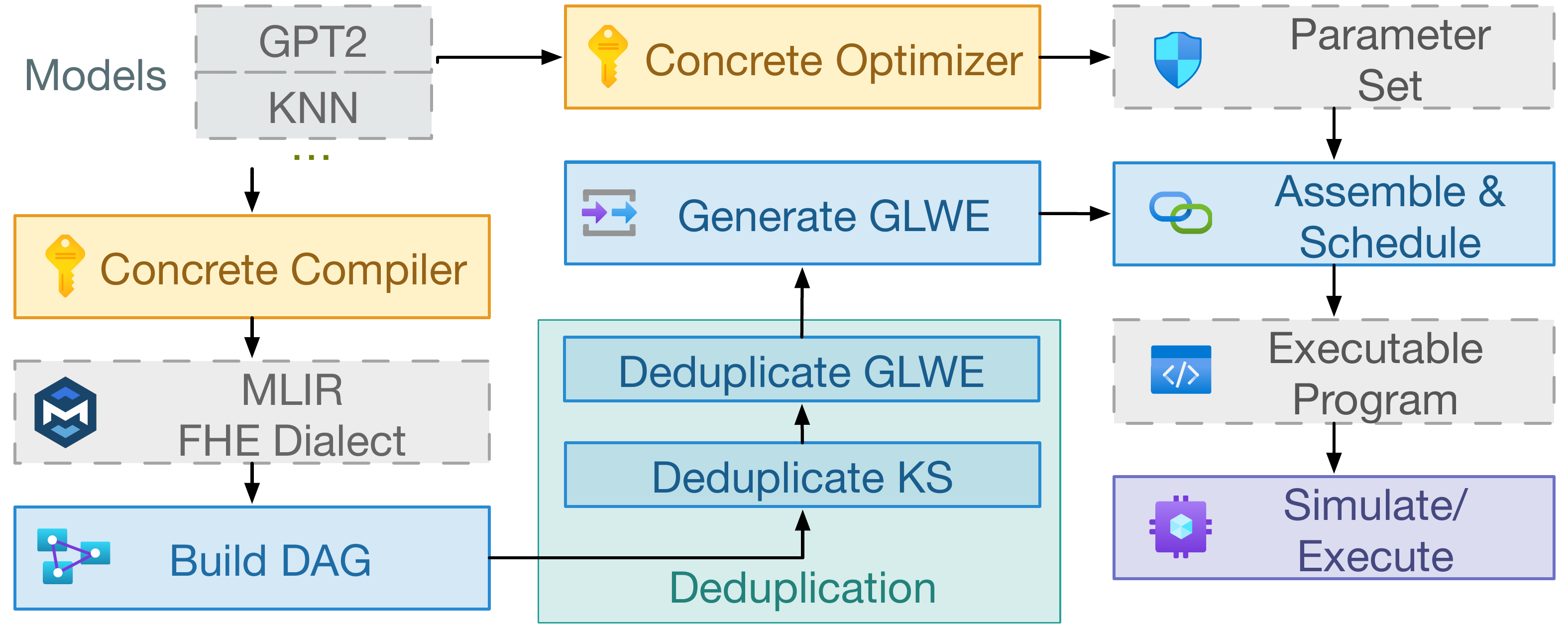}
   \caption{Compilation Workflow for Taurus.}
   \label{fig:compile}
\end{figure}

Figure~\ref{fig:compile} outlines our modular compilation framework, which integrates with the Concrete toolchain while maintaining flexibility through MLIR-based intermediate representations. We leverage Concrete's ecosystem for front-end processing and process programs in MLIR's \texttt{FHELinAlg} dialect~\cite{mlir}, ensuring Taurus supports any application expressible in this dialect.

We identified two deduplication optimizations: key-switching deduplication (\textit{KS-dedup}) and GLWE accumulator deduplication (\textit{ACC-dedup}).
KS-dedup enables reuse of key-switching results as inputs for multiple subsequent blind rotations when fanout structures exist. Unlike Boolean TFHE, which performs blind rotation first, Taurus performs key-switching first to enable this optimization. Multi-bit TFHE programs commonly apply multiple different LUTs to the same ciphertext, allowing KS-dedup to broadcast key-switching results to multiple BRUs. This reduces key-switching operations by up to 47.12\% in our evaluated workloads.

\paraobserv{Observation 6: Moving key-switching before blind rotation in PBS and treating PBS as a non-atomic operation create opportunities for deduplication in real-world TFHE workflows.}

ACC-dedup leverages the pattern where multi-bit TFHE programs frequently apply the same accumulator across multiple tensor elements. By sharing accumulators among tensor elements, this optimization reduces GLWE storage requirements by 91.54\%, significantly decreasing program sizes and DRAM capacity requirements.

\section{Evaluation}

\subsection{Design Space Exploration}
\textbf{Off-chip Memory Bandwidth Sensitivity}

\begin{figure}
    \centering
    \includegraphics[width=\linewidth]{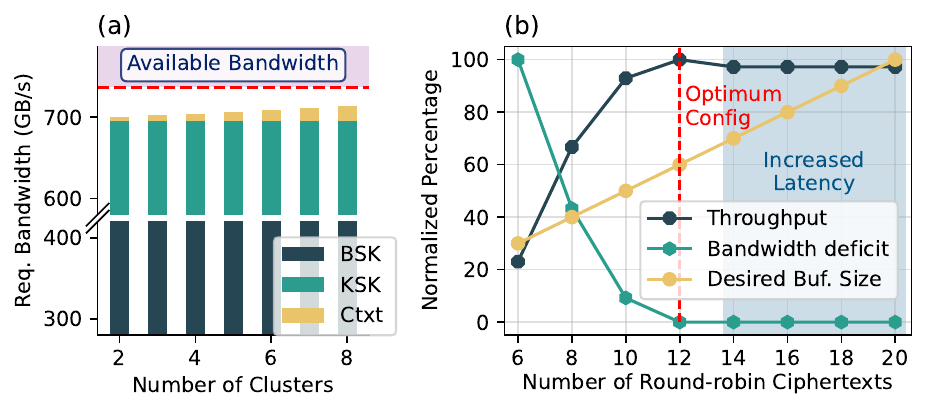}
    \caption{Architectural Analysis: (a) The impact of the number of clusters on the required bandwidth. The maximum bandwidth of two HBM2E stacks can satisfy the requirements of more than 8 clusters. (b) The impact of the number of round-robin ciphertexts on throughput, bandwidth deficits, and buffer sizes. Having 12 round-robin ciphertexts achieves maximum throughput while having the smallest buffer size.}
    \label{fig:dram_bandwidth}
\end{figure}

We analyze two key architectural parameters: the number of computation clusters and the number of round-robin ciphertexts per batch (48 ciphertexts scheduled simultaneously).

Figure~\ref{fig:dram_bandwidth}(a) shows that increasing clusters from 2 to 8 linearly increases bandwidth requirements for GLWE and LWE ciphertexts, while BSK and KSK bandwidth remains constant as keys are shared across clusters. Two HBM2E stacks provide sufficient bandwidth for up to 8 clusters.

Figure~\ref{fig:dram_bandwidth}(b) demonstrates that 12 round-robin ciphertexts achieve optimal performance by eliminating bandwidth deficits while minimizing buffer requirements. Beyond 12 ciphertexts, throughput plateaus while buffer size continues growing linearly, and excessive ciphertexts can cause underutilization when insufficient parallel operations exist.

\textbf{Impact of Buffer Size on Runtime}

\begin{figure}
    \centering
    \includegraphics[width=\linewidth]{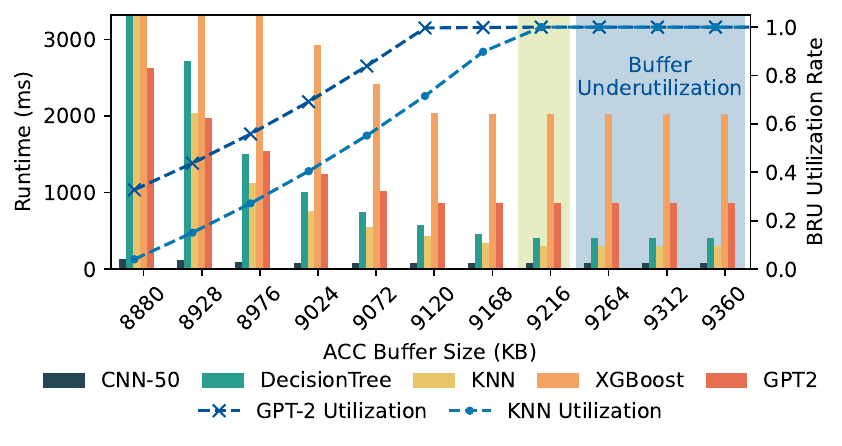}
    \caption{Impact of Accumulator Buffer Size on Wall-clock Runtime of Various Workloads and Utilization Rates}
    \label{fig:sram}
\end{figure}

The accumulator buffer is Taurus's largest buffer with a default size of $9216~\texttt{KB}$ to accommodate two GLWE accumulators for each ciphertext.

Figure~\ref{fig:sram} shows that reducing buffer size below $9216~\texttt{KB}$ forces data swapping to DRAM, stalling the BRU pipeline. The runtime point at which two HBM stacks cannot meet bandwidth requirements varies by polynomial degree. Utilization rates remain above 99\% in the $9120$--$9168~\texttt{KB}$ range, indicating relatively small swapping penalties.

Increasing buffer size beyond $9216~\texttt{KB}$ shows no utilization improvements but leads to underutilization unless round-robin ciphertexts are also increased.

\subsection{Hardware Implementation}

We implemented the Taurus architecture in Chisel HDL~\cite{chisel}, compiled into Verilog HDL, and synthesized by Synopsys Design Compiler~\cite{SynopsysDC} using TSMC \texttt{N16} process.
We modeled all scratchpad memories with the Arm Artisan physical IP compiler~\cite{ArmArtisan} and obtained all activity factors based on the worst-case parameter set.
We used DRAMSim3~\cite{dramsim3} to simulate two HBM stacks.
We pipelined the NoC and all functional units to achieve the design target of 1~GHz.
Table~\ref{tab:power_area} shows the total area and power consumption of Taurus, with breakdowns by major component.

\begin{table}
    \centering
    \caption{Area and Power Consumption Breakdown of Taurus}
    \begin{footnotesize}
    \label{tab:power_area}
    \begin{tabular}{lccr}
        \toprule
        Component & Area ($mm^2$) & Power ($W$)\\
        \hline
        Decomposer & 0.24 & 0.65\\
        2$\times$ FFT-A & 1.57 & 2.95\\
        FFT-B & 1.88 & 4.12\\
        VecMAC & 4.27 & 8.41\\
        Rotator & 0.18 & 0.63\\
        Transpose & 2.20 & 7.16\\
        VecMult & 2.06 & 4.06\\
        ModSwitch & $<$0.01 & $<$0.01\\
        \midrule
        BRU & 12.41 & 28.01\\
        LPU & 1.32 & 0.61\\
        I-FFT & 5.65 & 18.30\\
        \midrule
        Acc buf. (9.2MB) & 9.83 & 3.11\\
        GLWE buf. (1.5MB) & 1.88 & 0.52\\
        LWE buf. (24KB) & 0.02 & $<$0.01\\
        \midrule
        Cluster Group & 56.62 & 82.81\\
        GGSW buf. (0.8MB) & 1.22 & 0.91\\
        KSK buf. (0.5MB) & 0.50 & 0.07\\
        Twiddle buf. (0.8MB) & 1.39 & 0.27\\
        NoC & 0.16 & 0.43\\
        \midrule
        Total & 116.52 & 167.30\\
        \bottomrule
    \end{tabular}
    \end{footnotesize}
\end{table}

\subsection{Performance Evaluation}

\subsubsection{Real-world Workloads}
\label{sec:workloads}

{\footnotesize
    \begin{table}[t]
    \centering
    \caption{Wall-clock execution time comparison}
    \label{tab:runtime_comparison}
    \begin{footnotesize}
    \begin{tabular}{lccc }
        \toprule
        \specialleftcell{\textbf{Workload} \\ $n, (N, k)$, Width}&\specialcell{CPU (\texttt{s}) \\ w/ GPUs (\texttt{s})}&\ Taurus (\texttt{ms})&Speedup \\
        \hline
        \specialleftcell{\textbf{CNN-20 (PTQ)} \\ $737, (2048, 1)$, 6} & \specialcell{3.85 \\ 6.096} & 11.60 & \specialcell{331$\times$ \\ 525$\times$} \\
        \\[-1em]
        \specialleftcell{\textbf{CNN-50 (PTQ)} \\ $828, (4096, 1)$, 6} & \specialcell{15.31 \\ 49.714} & 74.27 & \specialcell{206$\times$ \\ 669$\times$} \\
        \\[-1em]
        \specialleftcell{\textbf{Decision Tree} \\ $1070, (65536, 1)$, 9} & \specialcell{645.40 \\ 522.2351} & 409.19 & \specialcell{1577$\times$ \\ 1276$\times$} \\
        \\[-1em]
        \specialleftcell{\textbf{GPT2} \\ $1003, (32768, 1)$, 6} & \specialcell{1218.13 \\ 721.14} & 860.94 & \specialcell{1414$\times$ \\ 837$\times$} \\
        \\[-1em]
        \specialleftcell{\textbf{GPT2 (12-head)} \\ $1009, (32768, 1)$, 6} & \specialcell{23685.14 \\ \texttt{OOM}} & 10649.33 & \specialcell{2224$\times$ \\ -} \\
        \\[-1em]
        \specialleftcell{\textbf{KNN} \\ $1058, (65536, 1)$, 9} & \specialcell{284.69 \\ 204.6} & 306.66 & \specialcell{928$\times$ \\ 667$\times$} \\
        \\[-1em]
        \specialleftcell{\textbf{XGBoost Reg} \\ $1025, (32768, 1)$, 8} & \specialcell{1793.27 \\ 912.11} & 689.29 & \specialcell{2601$\times$ \\ 1323$\times$} \\
        \\[-1em]
        \bottomrule
    \end{tabular}
    \end{footnotesize}
\end{table}

}

We evaluated Taurus using a cycle-accurate simulator across diverse Concrete toolchain workloads.

Our two-stage approach combines functionality modeling for computational correctness verification (matching Taurus hardware bit-by-bit) with performance modeling for cycle-accurate timing within seconds.
We use the state-of-the-art Concrete toolchain\footnote{Concrete-ML~\cite{ConcreteML} v1.6.1, Commit \texttt{8681124}. Concrete Compiler~\cite{Concrete} v2.7.0, Commit \texttt{b7793aeb}.} for fair CPU and GPU comparison.

Our evaluation spans low-precision workloads (two CNN models with 20 and 50 layers~\cite{concrete-ml-cnn}) and high-precision workloads including KNN classifier\footnote{Scikit-learn based~\cite{scikit-learn}, 3 neighbors, 30 leaves.}, XGBoost regressor\footnote{50 estimators, max depth 4, predicting Ames Housing prices~\cite{ames} with 6-bit quantization.}, decision tree classifier\footnote{Scikit-learn based~\cite{scikit-learn}, classifying Bioresponse dataset~\cite{bioresponse}, 18 max depth, 91 nodes, 7-bit quantization.}, and quantized GPT-2 models\footnote{HuggingFace pre-trained~\cite{gpt2}, 7-bit quantization, 6-bit rounding, single and 12-headed versions.}. Parameter sets are detailed in Table~\ref{tab:runtime_comparison}.

\subsubsection{Real-world Workloads Result Analysis}

Performance comparisons used an AMD EPYC 7R13 system (48 Zen 3 cores at 3.4~GHz, 256~GB DDR4-3200) with dual NVIDIA RTX A5000 GPUs. The 48 CPU cores match Taurus's parallel ciphertext processing capacity.

\textbf{Low bit-width Workloads}
CNN models achieve up to $277\times$ speedup with single-ciphertext bootstrapping latencies of 0.28~ms (CNN-20) and 0.85~ms (CNN-50).
Compared to Morphling, CNN-20 improves from 0.34~s to 0.0139~s and CNN-50 from 1.72~s to 0.0925~s through compiler optimizations and enhanced polynomial multiplication throughput.

\textbf{High bit-width Workloads}
High bit-width workloads show dramatic improvements up to $2595\times$ over CPU and $1320\times$ over GPU, with single-ciphertext bootstrapping latencies ranging from 6.16~ms to 34.67~ms. XGBoost achieved the highest utilization through highly parallel LUT evaluations. This work demonstrates the first homomorphically encrypted GPT-2 inference at usable speeds.

\textbf{Utilization and Batchsize}

\begin{figure}
    \centering
    \includegraphics[width=\linewidth]{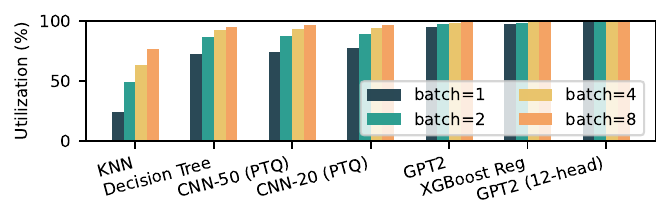}
    \caption{Taurus Cluster Utilization vs. Workloads}
    \label{fig:utilization}
\end{figure}

Real-world deployments typically process multiple queries in batches~\cite{batch_cnn, quantize_inference}.
Figure~\ref{fig:utilization} shows how input batch size affects hardware utilization across different workloads.
For predominantly serial workloads like KNN and Decision Tree, larger batch sizes substantially improve utilization, with KNN reaching 75\% utilization at batch size 8.

\paraobserv{Observation 7: The parallelism offered by TFHE hardware accelerators can be effectively harnessed in real-world multi-bit TFHE workloads as batch sizes increase.}

\subsection{Quantifying Memory Bandwidth and Architectural Contributions}

\begin{figure}
    \centering
    \includegraphics[width=\linewidth]{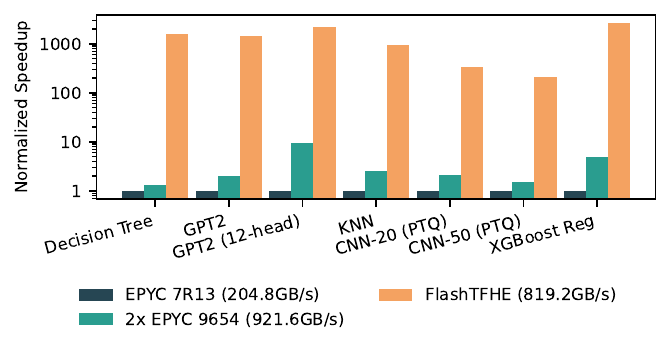}
    \caption{Normalized Speedup Comparison Across Platforms (Log Scale)}
    \label{fig:norm-speedup}
\end{figure}

We measured workload performance on dual AMD EPYC 9654 CPUs with 921.6~GB/s total memory bandwidth, exceeding Taurus's two HBM stacks (819~GB/s).

Figure~\ref{fig:norm-speedup} compares normalized speedup of dual EPYC 9654 and Taurus relative to EPYC 7R13 (baseline).
The dual 9654 improvement stems from $4\times$ more cores (192 vs 48), $4.5\times$ memory bandwidth increase, up to 13\% IPC boost, AVX-512 support, and higher TDP (800~W vs 270~W).
Since dual 9654 provides superior memory bandwidth, the additional Taurus improvement primarily demonstrates microarchitectural contributions.

\subsection{Comparison with Prior TFHE Accelerators}

\small
\begin{table}
    \centering
    \footnotesize
    \caption{ASIC Area Comparison Table}
    \label{tab:asic_comparison}
    \begin{tabular}{lccc}
        \toprule
        \textbf{Accelerator} & \specialcell{Reported \\ Area ($mm^2$)} & \specialcell{Area in \\ 16nm ($mm^2$)} & \specialcell{PolyMult \\/Unit Area} \\ 
        \hline
        \textbf{Strix} & 141.37 & 52.69 & 1.21 \\
        \\[-1em]
        \textbf{MATCHA} & 36.96 & 25.08 & 1.27 \\
        \\[-1em]
        \textbf{Morphling} & 74.79 & 24.95 & 10.25 \\
        \\[-1em]
        \textbf{Taurus} & 116.52 & 116.52 & 17.58 \\
        \\[-1em]
        \bottomrule
    \end{tabular}
    \normalsize
\end{table}

\normalsize

\textbf{Throughput per unit area comparison.}

Table~\ref{tab:asic_comparison} compares polynomial multiplication throughput with existing ASIC accelerators~\cite{matcha, Strix, morphling}, with areas scaled to 16~nm~\cite{scaling}.\footnote{Throughputs measured at $k=1$ for typical multi-bit TFHE workloads. Memory controllers and PHY excluded for fair comparison.} 

Taurus achieves slightly better PolyMult throughput per unit area while supporting much higher polynomial degrees ($2^{16}$ vs Morphling's 4096), enabling 10-bit multi-bit TFHE programs compared to the previous 5-bit limitation.
Expanding complex number width from 32 to 48 bits provides plug-and-play compatibility with industrial toolchains like Concrete~\cite{Concrete, ConcreteML}.

\textbf{Runtime comparison on benchmarks with XPU-like design.}

\begin{table}[t]
    \centering
    \caption{Runtime on Taurus and Taurus with extended XPU}
    \label{tab:morphling_runtime}
    \begin{footnotesize}
    \begin{tabular}{lccc }
        \toprule
        \textbf{Workload} & \specialleftcell{$\text{Taurus}_{\text{XPU}}$ \\ (\texttt{ms})} & \specialleftcell{Taurus \\ (\texttt{ms})} & Speedup \\
        \hline
        \textbf{CNN-20 (PTQ)} & 78.65 & 11.60 & 6.78$\times$ \\
        \\[-1em]
        \textbf{CNN-50 (PTQ)} & 506.27 & 74.27 & 6.82$\times$ \\
        \\[-1em]
        \textbf{Decision Tree} & 2794.60 & 409.19 & 6.83$\times$ \\
        \\[-1em]
        \textbf{GPT2} & 5851.00 & 860.94 & 6.80$\times$ \\
        \\[-1em]
        \textbf{GPT2 (12-head)} & 75219.27 & 10649.33 & 7.06$\times$ \\
        \\[-1em]
        \textbf{KNN} & 982.49 & 306.66 & 3.20$\times$ \\
        \\[-1em]
        \textbf{XGBoost Reg} & 4749.30 & 689.29 & 6.89$\times$ \\
        \\[-1em]
        \bottomrule
    \end{tabular}
    \end{footnotesize}
\end{table}

To evaluate Taurus's architectural advantages over Morphling's systolic array design (state-of-the-art), we implemented a variant that replaces our BRU with Morphling's XPU for external product operations.
We extended the R2MDC FFT units in the XPU to support higher polynomial degrees required by our workloads.
Table~\ref{tab:morphling_runtime} compares the runtime performance between Taurus and this XPU-based variant (denoted as $\text{Taurus}_{\text{XPU}}$).
Taurus achieves a consistent 3-7$\times$ speedup across all benchmarks, with most workloads showing approximately 6.8$\times$ improvement.
These results demonstrate the effectiveness of our round-robin scheduling and high-throughput FFT design compared to Morphling's systolic array approach.

\section{Conclusions}

We presented Taurus, a hardware architecture that makes multi-bit TFHE practical at wider numeric widths and scaled parameter sets.
Taurus supports up to 10-bit ciphertexts at 128-bit security, and delivers up to $2600\times$ and $1200\times$ speedups over CPU and GPU baselines, respectively.
Compare to a Morphling-style XPU variant scaled to larger sizes, it shows $3$--$7\times$ faster, while offering higher poly-multiplication throughput per unit area than previous designs.
We also pair the architecture with a companion compiler that applies deduplications to expose more reuse, further improving end-to-end efficiency.



\bibliographystyle{IEEEtranS}
\bibliography{refs}

\begin{thebibliography}{10}
\providecommand{\url}[1]{#1}
\csname url@samestyle\endcsname
\providecommand{\newblock}{\relax}
\providecommand{\bibinfo}[2]{#2}
\providecommand{\BIBentrySTDinterwordspacing}{\spaceskip=0pt\relax}
\providecommand{\BIBentryALTinterwordstretchfactor}{4}
\providecommand{\BIBentryALTinterwordspacing}{\spaceskip=\fontdimen2\font plus
\BIBentryALTinterwordstretchfactor\fontdimen3\font minus
  \fontdimen4\font\relax}
\providecommand{\BIBforeignlanguage}[2]{{%
\expandafter\ifx\csname l@#1\endcsname\relax
\typeout{** WARNING: IEEEtranS.bst: No hyphenation pattern has been}%
\typeout{** loaded for the language `#1'. Using the pattern for}%
\typeout{** the default language instead.}%
\else
\language=\csname l@#1\endcsname
\fi
#2}}
\providecommand{\BIBdecl}{\relax}
\BIBdecl

\bibitem{bfv}
A.~Al~Badawi, Y.~Polyakov, K.~M.~M. Aung, B.~Veeravalli, and K.~Rohloff,
  ``Implementation and {{Performance Evaluation}} of {{RNS Variants}} of the
  {{BFV Homomorphic Encryption Scheme}},'' \emph{IEEE Transactions on Emerging
  Topics in Computing}, vol.~9, no.~2, pp. 941--956, Apr. 2021.

\bibitem{lwe_estimator}
\BIBentryALTinterwordspacing
M.~R. Albrecht, R.~Player, and S.~Scott, ``\BIBforeignlanguage{en}{On the
  concrete hardness of {Learning} with {Errors}},''
  \emph{\BIBforeignlanguage{en}{Journal of Mathematical Cryptology}}, vol.~9,
  no.~3, pp. 169--203, Oct. 2015. [Online]. Available:
  \url{https://www.degruyter.com/document/doi/10.1515/jmc-2015-0016/html}
\BIBentrySTDinterwordspacing

\bibitem{ArmArtisan}
{Arm Limited}, ``Arm {{Artisan IP}}: {{Boost SoC Design Efficiency}},''
  \url{https://www.arm.com/products/silicon-ip-physical/artisan-ip}, 2024.

\bibitem{chisel}
J.~Bachrach, H.~Vo, B.~Richards, Y.~Lee, A.~Waterman, R.~Avi{\v z}ienis,
  J.~Wawrzynek, and K.~Asanovi{\'c}, ``Chisel: Constructing hardware in a
  {{Scala}} embedded language,'' in \emph{Proceedings of the 49th {{Annual
  Design Automation Conference}}}.\hskip 1em plus 0.5em minus 0.4em\relax San
  Francisco California: ACM, Jun. 2012, pp. 1216--1225.

\bibitem{ks_first}
F.~Bourse, M.~Minelli, M.~Minihold, and P.~Paillier, ``Fast {{Homomorphic
  Evaluation}} of {{Deep Discretized Neural Networks}},'' in \emph{Advances in
  {{Cryptology}} -- {{CRYPTO}} 2018}, H.~Shacham and A.~Boldyreva, Eds.\hskip
  1em plus 0.5em minus 0.4em\relax Cham: Springer International Publishing,
  2018, vol. 10993, pp. 483--512.

\bibitem{rns-ckks}
J.~H. Cheon, K.~Han, A.~Kim, M.~Kim, and Y.~Song, ``A {{Full RNS Variant}} of
  {{Approximate Homomorphic Encryption}},'' in \emph{Selected {{Areas}} in
  {{Cryptography}} -- {{SAC}} 2018}, C.~Cid and M.~J. Jacobson, Eds.\hskip 1em
  plus 0.5em minus 0.4em\relax Cham: Springer International Publishing, 2019,
  vol. 11349, pp. 347--368.

\bibitem{concrete-ml-cnn}
B.~{Chevallier-Mames} and K.~Celia, ``Making {{FHE Faster}} for {{ML}}:
  {{Beating}} our {{Previous Paper Benchmarks}} with {{Concrete ML}},''
  \url{https://www.zama.ai/post/making-fhe-faster-for-ml-beating-our-previous-paper-benchmarks-with-concrete-ml},
  Jul. 2024.

\bibitem{chillotti_tfhe_2020}
I.~Chillotti, N.~Gama, M.~Georgieva, and M.~Izabach{\`e}ne, ``{{TFHE}}: {{Fast
  Fully Homomorphic Encryption Over}} the {{Torus}},'' \emph{Journal of
  Cryptology}, vol.~33, no.~1, pp. 34--91, Jan. 2020.

\bibitem{concrete-cnn}
I.~Chillotti, M.~Joye, and P.~Paillier, ``Programmable {{Bootstrapping Enables
  Efficient Homomorphic Inference}} of {{Deep Neural Networks}},'' in
  \emph{Cyber {{Security Cryptography}} and {{Machine Learning}}}, S.~Dolev,
  O.~Margalit, B.~Pinkas, and A.~Schwarzmann, Eds.\hskip 1em plus 0.5em minus
  0.4em\relax Cham: Springer International Publishing, 2021, vol. 12716, pp.
  1--19.

\bibitem{wop-pbs}
I.~Chillotti, D.~Ligier, J.-B. Orfila, and S.~Tap, ``Improved {{Programmable
  Bootstrapping}} with {{Larger Precision}} and {{Efficient Arithmetic
  Circuits}} for {{TFHE}},'' in \emph{Advances in {{Cryptology}} --
  {{ASIACRYPT}} 2021}, M.~Tibouchi and H.~Wang, Eds.\hskip 1em plus 0.5em minus
  0.4em\relax Cham: Springer International Publishing, 2021, vol. 13092, pp.
  670--699.

\bibitem{ames}
D.~De~Cock, ``Ames, {{Iowa}}: {{Alternative}} to the {{Boston Housing Data}} as
  an {{End}} of {{Semester Regression Project}},'' \emph{Journal of Statistics
  Education}, vol.~19, no.~3, p.~8, Nov. 2011.

\bibitem{fhew}
L.~Ducas and D.~Micciancio, ``{{FHEW}}: {{Bootstrapping Homomorphic
  Encryption}} in {{Less Than}} a {{Second}},'' in \emph{Advances in
  {{Cryptology}} -- {{EUROCRYPT}} 2015}, E.~Oswald and M.~Fischlin, Eds.\hskip
  1em plus 0.5em minus 0.4em\relax Berlin, Heidelberg: Springer Berlin
  Heidelberg, 2015, vol. 9056, pp. 617--640.

\bibitem{first_relu}
K.~Fukushima, ``Neocognitron: {{A}} self-organizing neural network model for a
  mechanism of pattern recognition unaffected by shift in position,''
  \emph{Biological Cybernetics}, vol.~36, no.~4, pp. 193--202, Apr. 1980.

\bibitem{gsw}
C.~Gentry, A.~Sahai, and B.~Waters, ``Homomorphic {{Encryption}} from
  {{Learning}} with {{Errors}}: {{Conceptually-Simpler}},
  {{Asymptotically-Faster}}, {{Attribute-Based}},'' in \emph{Advances in
  {{Cryptology}} -- {{CRYPTO}} 2013}, R.~Canetti and J.~A. Garay, Eds.\hskip
  1em plus 0.5em minus 0.4em\relax Berlin, Heidelberg: Springer Berlin
  Heidelberg, 2013, vol. 8042, pp. 75--92.

\bibitem{double-real2}
A.~X. Glittas, M.~Sellathurai, and G.~Lakshminarayanan, ``A {{Normal I}}/{{O
  Order Radix-2 FFT Architecture}} to {{Process Twin Data Streams}} for
  {{MIMO}},'' \emph{IEEE Transactions on Very Large Scale Integration (VLSI)
  Systems}, vol.~24, no.~6, pp. 2402--2406, Jun. 2016.

\bibitem{bioresponse}
B.~Hamner, {dcthompson}, and {Jorg}, ``Predicting a biological response,''
  \url{https://kaggle.com/competitions/bioresponse}, 2012.

\bibitem{ckks_regression}
K.~Han, S.~Hong, J.~H. Cheon, and D.~Park, ``Logistic {{Regression}} on
  {{Homomorphic Encrypted Data}} at {{Scale}},'' \emph{Proceedings of the AAAI
  Conference on Artificial Intelligence}, vol.~33, no.~01, pp. 9466--9471, Jul.
  2019.

\bibitem{matcha}
L.~Jiang, Q.~Lou, and N.~Joshi, ``{{MATCHA}}: A fast and energy-efficient
  accelerator for fully homomorphic encryption over the torus,'' in
  \emph{Proceedings of the 59th {{ACM}}/{{IEEE Design Automation
  Conference}}}.\hskip 1em plus 0.5em minus 0.4em\relax San Francisco
  California: ACM, Jul. 2022, pp. 235--240.

\bibitem{Marc_guide}
M.~Joye, ``Guide to {{Fully Homomorphic Encryption}} over the [{{Discretized}}]
  {{Torus}},'' \url{https://eprint.iacr.org/2021/1402}, Oct. 2021.

\bibitem{sharp}
J.~Kim, S.~Kim, J.~Choi, J.~Park, D.~Kim, and J.~H. Ahn, ``{{SHARP}}: {{A
  Short-Word Hierarchical Accelerator}} for {{Robust}} and {{Practical Fully
  Homomorphic Encryption}},'' in \emph{Proceedings of the 50th {{Annual
  International Symposium}} on {{Computer Architecture}}}.\hskip 1em plus 0.5em
  minus 0.4em\relax Orlando FL USA: ACM, Jun. 2023, pp. 1--15.

\bibitem{ark}
J.~Kim, G.~Lee, S.~Kim, G.~Sohn, M.~Rhu, J.~Kim, and J.~H. Ahn, ``{{ARK}}:
  {{Fully Homomorphic Encryption Accelerator}} with {{Runtime Data Generation}}
  and {{Inter-Operation Key Reuse}},'' in \emph{2022 55th {{IEEE}}/{{ACM
  International Symposium}} on {{Microarchitecture}} ({{MICRO}})}.\hskip 1em
  plus 0.5em minus 0.4em\relax Chicago, IL, USA: IEEE, Oct. 2022, pp.
  1237--1254.

\bibitem{bts}
S.~Kim, J.~Kim, M.~J. Kim, W.~Jung, J.~Kim, M.~Rhu, and J.~H. Ahn, ``{{BTS}}:
  An accelerator for bootstrappable fully homomorphic encryption,'' in
  \emph{Proceedings of the 49th {{Annual International Symposium}} on
  {{Computer Architecture}}}.\hskip 1em plus 0.5em minus 0.4em\relax New York
  New York: ACM, Jun. 2022, pp. 711--725.

\bibitem{Hitchhiker_guide}
J.~Klemsa, ``Hitchhiker's {{Guide}} to the {{TFHE Scheme}},''
  \url{https://eprint.iacr.org/2022/1315}, Oct. 2022.

\bibitem{batch_cnn}
\BIBentryALTinterwordspacing
J.~Kosaian, A.~Phanishayee, M.~Philipose, D.~Dey, and R.~Vinayak, ``Boosting
  the throughput and accelerator utilization of specialized cnn inference
  beyond increasing batch size,'' in \emph{Proceedings of the 38th
  International Conference on Machine Learning}, ser. Proceedings of Machine
  Learning Research, M.~Meila and T.~Zhang, Eds., vol. 139.\hskip 1em plus
  0.5em minus 0.4em\relax PMLR, 18--24 Jul 2021, pp. 5731--5741. [Online].
  Available: \url{https://proceedings.mlr.press/v139/kosaian21a.html}
\BIBentrySTDinterwordspacing

\bibitem{quantize_inference}
R.~Krishnamoorthi, ``Quantizing deep convolutional networks for efficient
  inference: {{A}} whitepaper,'' \url{http://arxiv.org/abs/1806.08342}, Jun.
  2018.

\bibitem{mlir}
C.~Lattner, M.~Amini, U.~Bondhugula, A.~Cohen, A.~Davis, J.~Pienaar, R.~Riddle,
  T.~Shpeisman, N.~Vasilache, and O.~Zinenko, ``{{MLIR}}: Scaling compiler
  infrastructure for domain specific computation,'' in \emph{2021 {{IEEE/ACM}}
  International Symposium on Code Generation and Optimization (CGO)}, 2021, pp.
  2--14.

\bibitem{dramsim3}
S.~Li, Z.~Yang, D.~Reddy, A.~Srivastava, and B.~Jacob, ``{{DRAMsim3}}: {{A
  Cycle-Accurate}}, {{Thermal-Capable DRAM Simulator}},'' \emph{IEEE Computer
  Architecture Letters}, vol.~19, no.~2, pp. 106--109, Jul. 2020.

\bibitem{PyTFHE}
J.~Ma, C.~Xu, and L.~W. Wills, ``{{PyTFHE}}: {{An End-to-End Compilation}} and
  {{Execution Framework}} for {{Fully Homomorphic Encryption Applications}},''
  in \emph{2023 {{IEEE International Symposium}} on {{Performance Analysis}} of
  {{Systems}} and {{Software}} ({{ISPASS}})}.\hskip 1em plus 0.5em minus
  0.4em\relax Raleigh, NC, USA: IEEE, Apr. 2023, pp. 24--34.

\bibitem{fhe_survey}
C.~Marcolla, V.~Sucasas, M.~Manzano, R.~Bassoli, F.~H.~P. Fitzek, and N.~Aaraj,
  ``Survey on {{Fully Homomorphic Encryption}}, {{Theory}}, and
  {{Applications}},'' \emph{Proceedings of the IEEE}, vol. 110, no.~10, pp.
  1572--1609, Oct. 2022.

\bibitem{virtual-riscv}
K.~Matsuoka, R.~Banno, N.~Matsumoto, T.~Sato, and S.~Bian, ``Virtual {{Secure
  Platform}}: A {{Five-Stage}} pipeline processor over {{TFHE}},'' in
  \emph{30th {{USENIX}} Security Symposium ({{USENIX}} Security 21)}.\hskip 1em
  plus 0.5em minus 0.4em\relax USENIX Association, Aug. 2021, pp. 4007--4024,
  \url{https://www.usenix.org/conference/usenixsecurity21/presentation/matsuoka}.

\bibitem{scikit-learn}
F.~Pedregosa, G.~Varoquaux, A.~Gramfort, V.~Michel, B.~Thirion, O.~Grisel,
  M.~Blondel, A.~M{\"u}ller, J.~Nothman, G.~Louppe, P.~Prettenhofer, R.~Weiss,
  V.~Dubourg, J.~Vanderplas, A.~Passos, D.~Cournapeau, M.~Brucher, M.~Perrot,
  and {\'E}.~Duchesnay, ``Scikit-learn: {{Machine}} learning in python,''
  \url{https://arxiv.org/abs/1201.0490}, Jan. 2012.

\bibitem{morphling}
{Prasetiyo}, A.~Putra, and J.-Y. Kim, ``Morphling: {{A Throughput-Maximized
  TFHE-based Accelerator}} using {{Transform-domain Reuse}},'' in \emph{2024
  {{IEEE International Symposium}} on {{High-Performance Computer
  Architecture}} ({{HPCA}})}.\hskip 1em plus 0.5em minus 0.4em\relax Edinburgh,
  United Kingdom: IEEE, Mar. 2024, pp. 249--262.

\bibitem{Strix}
A.~Putra, {Prasetiyo}, Y.~Chen, J.~Kim, and J.-Y. Kim, ``Strix: {{An}}
  end-to-end streaming architecture with two-level ciphertext batching for
  fully homomorphic encryption with programmable bootstrapping,'' in
  \emph{Proceedings of the 56th Annual {{IEEE}}/{{ACM}} International Symposium
  on Microarchitecture}, ser. Micro '23.\hskip 1em plus 0.5em minus 0.4em\relax
  New York, NY, USA: ACM, 2023, pp. 1319--1331.

\bibitem{gpt2}
A.~Radford, J.~Wu, R.~Child, D.~Luan, D.~Amodei, and I.~Sutskever, ``Language
  {{Models}} are {{Unsupervised Multitask Learners}},'' \emph{OpenAI blog},
  vol.~1, no.~8, p.~9, 2019.

\bibitem{f1}
N.~Samardzic, A.~Feldmann, A.~Krastev, S.~Devadas, R.~Dreslinski, C.~Peikert,
  and D.~Sanchez, ``F1: A fast and programmable accelerator for fully
  homomorphic encryption,'' in \emph{{{MICRO-54}}: 54th Annual {{IEEE}}/{{ACM}}
  International Symposium on Microarchitecture}, ser. Micro '21.\hskip 1em plus
  0.5em minus 0.4em\relax New York, NY, USA: Association for Computing
  Machinery, 2021, pp. 238--252.

\bibitem{craterlake}
N.~Samardzic, A.~Feldmann, A.~Krastev, N.~Manohar, N.~Genise, S.~Devadas,
  K.~Eldefrawy, C.~Peikert, and D.~Sanchez, ``{{CraterLake}}: A hardware
  accelerator for efficient unbounded computation on encrypted data,'' in
  \emph{Proceedings of the 49th {{Annual International Symposium}} on
  {{Computer Architecture}}}.\hskip 1em plus 0.5em minus 0.4em\relax New York
  New York: ACM, Jun. 2022, pp. 173--187.

\bibitem{scaling}
A.~Stillmaker and B.~Baas, ``Scaling equations for the accurate prediction of
  {{CMOS}} device performance from 180 nm to 7 nm,'' \emph{Integration},
  vol.~58, pp. 74--81, Jun. 2017.

\bibitem{SynopsysDC}
{Synopsys, Inc.}, ``Design {{Compiler}}: {{RTL Synthesis Solution}},''
  \url{https://www.synopsys.com/implementation-and-signoff/rtl-synthesis-test/dc-ultra.html},
  Mountain View, CA, USA, Sep. 2024.

\bibitem{ConcreteML}
{Zama}, ``Concrete {{ML}}: A privacy-preserving machine learning library using
  fully homomorphic encryption for data scientists,''
  \url{https://github.com/zama-ai/concrete-ml}, 2022.

\bibitem{Concrete}
{Zama}, ``Concrete: {{TFHE Compiler}} that converts {{Python}} programs into
  {{FHE}} equivalent,'' \url{https://github.com/zama-ai/concrete}, 2022.

\end{thebibliography}

\end{document}